\documentclass[preprint]{aastex}

\usepackage{amsmath, tikz}

\newcommand{\tens}[1]{\ensuremath{\mathcal{#1}}}
\renewcommand{\vec}[1]{\ensuremath{\mathbf{#1}}}

\definecolor{lightgrey}{gray}{0.7}
\definecolor{darkgrey}{gray}{0.3}
\definecolor{mygrey}{gray}{0.4}
\definecolor{insbOrange}{rgb}{0.945, 0.6, 0.0784}

\shorttitle{The CRONOS code for astrophysical MHD}
\shortauthors{Kissmann et al.}

\begin{document}


\title{Colliding-wind Binaries with strong magnetic fields}


\author{R. Kissmann and K. Reitberger and O. Reimer}
\affil{Institut f\"ur Astro- und Teilchenphysik, Universit\"at Innsbruck, Austria}
\and
\author{A. Reimer and E. Grimaldo}
\affil{Institut f\"ur theoretische Physik, Universit\"at Innsbruck, Austria}
\email{ralf.kissmann@uibk.ac.at}




\begin{abstract}
The dynamics of colliding wind binary systems and conditions for efficient particle acceleration therein have attracted multiple numerical studies in the recent years. These numerical models seek an explanation of the thermal and non-thermal emission of these systems as seen by observations. In the non-thermal regime, radio and X-ray emission is observed for several of these colliding-wind binaries, while gamma-ray emission has so far only been found in $\eta$ Carinae and possibly in WR 11. Energetic electrons are deemed responsible for a large fraction of the observed high-energy photons in these systems. Only in the gamma-ray regime there might be, depending on the properties of the stars, a significant contribution of emission from neutral pion decay. Thus, studying the emission from colliding-wind binaries requires detailed models of the acceleration and propagation of energetic electrons. This in turn requires a detailed understanding of the magnetic field, which will not only affect the energy losses of the electrons but in case of synchrotron emission also the directional dependence of the emissivity.
In this study we investigate magnetohydrodynamic simulations of different colliding wind binary systems with magnetic fields that are strong enough to have a significant effect on the winds. Such strong fields require a detailed treatment of the near-star wind acceleration zone. We show the implementation of such simulations and discuss  results that demonstrate the effect of the magnetic field on the structure of the wind collision region.

\end{abstract}





\section{Introduction}
Massive, luminous stars of spectral type O and B and also Wolf-Rayet (WR) stars are known to drive mass-loaded and fast stellar winds. In systems containing two such luminous stars the supersonic winds interact with each other, thereby forming a wind-collision region (WCR) enclosed by two strong shock waves \citep[see, e.g.][]{Usov1992ApJ389_635}. These shocks have the potential to accelerate particles to sufficiently high energies that the WCR becomes visible in non-thermal radiation \citep[see, e.g.,][]{EichlerUsov1993ApJ402_271,DoughertyWilliams2000MNRAS319_1005, DoughertyEtAl2003AnA409_217, NiemelaEtAl1998AJ115_2047}. The relevant emission channels of non-thermal radiation are synchrotron emission in the radio regime \citep[see, e.g.,][]{DoughertyEtAl2005ApJ623_447,WilliamsEtAl1997MNRAS289_10,Pittard2010MNRAS403_1633}, inverse Compton emission in the X-ray regime \citep[see, e.g.][]{PittardEtAl2010MNRAS403_1657} and both inverse Compton and pion decay processes at higher energies \citep[see, e.g.][]{BenagliaRomero2003AnA399_1121,ReimerEtAl2006ApJ644_1118,ReitbergerEtAl2014ApJ789_87}.
For an overview of observations of colliding-wind binaries (CWBs) at different energies see \citet{DeBecker2013AnA558A_28}. 


Apparently, a large fraction of the non-thermal photons result from energy-loss processes of electrons accelerated at the shock fronts. The acceleration and energy losses of these electrons strongly depend on the strength and direction of the magnetic field in and near the WCR. The majority of the recent modeling efforts, however, was based on hydrodynamical modeling of the interacting stellar winds. In these cases, the magnetic field was prescribed analytically: \citet{DoughertyEtAl2003AnA409_217,PittardDougherty2006MNRAS372_801} assumed the magnetic energy density to be proportional to the thermal pressure, while \citet{ReitbergerEtAl2014ApJ782_96,ReitbergerEtAl2014ApJ789_87} used an analytical description for the magnetic field from each individual star. Such prescriptions, however, are rather simplified. Apart from that, no direction of the magnetic field can be inferred from these approaches. Thus, neither an acceleration efficiency depending on the obliquity of the shocks nor a consistent treatment of the polarization of synchrotron emission can be taken into account.

\citet{FalcetaGoncalvesAbraham2012MNRAS423_1562} introduced the first magnetohydrodynamical models of a CWB system. In these simulations, a dipole field with a polar field strength of $B \sim 10^{-4}$ Tesla for each star was prescribed, and the magnetic field in the stellar wind plasma was evolved consistently in time. Such a comparatively weak field is transported passively with the flow of the stellar winds without having a relevant back reaction on the wind evolution. Observations, however, indicate that magnetic fields of early type stars might be significantly stronger than that \citep[see, e.g.,][]{AuriereEtAl2007AnA475_1053,PetitEtAl2013MNRAS429_398,ChevrotiereEtAl2014ApJ781_73,
FossatiEtAl2015AnA574A_20,FossatiEtAl2015AnA582A_45,WadeEtAl2016MNRAS456_2}. However, there is also a wide range of stars for which only upper limits could be determined. With regard to stars in particle-accelerating CWB systems no such strong fields have been observed, yet. For several such systems \citet{NeinerEtAl2015AnA575A_66} find upper limits for the polar magnetic-field strength on the order of 0.02 Tesla.

This discussion shows that it is certainly worthwhile to investigate CWB systems with polar magnetic fields stronger than $\sim10^{-4}$ Tesla. In this case the change of the stellar wind outflow due to the presence of the magnetic field has to be taken into account. This is particularly important for the acceleration region of the wind near the stellar surface. Correspondingly, we investigate magnetohydrodynamical models of such systems with surface magnetic fields on the order of 0.01 Tesla. For this we introduce a multi-step procedure that assures the consistent treatment of magnetic field and wind acceleration near the stellar surface.  In the next section we present this method together with the description of the numerical model used in this study. In Sec. \ref{SecResults} we then discuss the results and implications of our study of a range of models of magnetized CWBs. Finally, we summarize our findings in Sec. \ref{SecConclusion}.



\section{Numerical Model}
Simulations of CWBs with stars having line-driven winds are numerically a demanding task, particularly for magnetized winds, because of the very steep velocity gradient near the stellar surface. Thus, the region around each star has to be studied in much greater detail, i.e. with higher resolution, than the more distant parts of the wind outflow.
 This resolution requirement is at odds with the WCR being the focus of an analysis of the dynamics of a CWB. In hydrodynamics these problems can be circumvented by using an adapted beta-law prescription \citep[see][]{LamersCassinelli1999Book} in the vicinity of the star \citep[see][]{ReitbergerEtAl2014ApJ782_96}. If magnetic fields are taken into account, however, they can have a significant influence on the wind-acceleration, predominantly near the stellar surface \citep[see][]{udDoulaOwocki2002ApJ576_413}. These authors also show that only a sufficiently small magnetic field has a negligible effect on the wind acceleration and can, thus, be treated as a passive tracer. Such a setup has been investigated by \citet{FalcetaGoncalvesAbraham2012MNRAS423_1562} for a CWB system with rather low ($10^{-4}$\,T) surface magnetic fields. In contrast to that we aim at including significantly higher surface magnetic fields, where the effects of the wind acceleration can not be disregarded anymore. Before we introduce our setup for the near-star resolution problem, we start by discussing the relevant system of equations.

\subsection{Mathematical Description}
We study stellar winds in CWB systems influenced by given stellar magnetic fields. Inside the stars the magnetic fields are assumed to be of dipole character and are kept constant throughout the simulations. Outside the stars the stellar winds are allowed to evolve with the magnetic field, where we use ideal magnetohydrodynamics (MHD) to describe the plasma of the stellar winds:
\begin{align}
  \label{EqContinuity}
  \frac{\partial \rho}{\partial t}
  &+
  \nabla \cdot \left(\rho \vec{u}\right)
  =
  0
  \\
  \label{EqMomConservation}
  \frac{\partial \left(\rho \vec{u} \right)}{\partial t}
  &+
  \nabla \cdot \left(\rho \vec{u} \vec{u} \right)
  + \nabla \cdot \left(p \tens{I}\right)
  + \frac{1}{\mu_0}\vec{B} \times \left(\nabla \times \vec{B}\right)
  =
  \vec{f}
  \\
  \label{EqInduction}
  \index{Induction equation}
  \frac{\partial \vec{B}}{\partial t}
  &=
  \nabla \times \left(\vec{u} \times \vec{B}\right)
  \\
  \label{EqEnergyConservation}
  \frac{\partial e}{\partial t}
  &
  + \nabla \cdot \left(
  \left(e + \frac{B^2}{2 \mu_0} + p \right)\vec{u}
  - \frac{1}{\mu_0}(\vec{u}\cdot\vec{B})\vec{B}
  \right)
  = \vec{u} \cdot \vec{f}
  +
  \left(\frac{\rho}{m_H}\right)^2 \Lambda(T)
\end{align}
Here, the dynamical variables are the density $\rho$, the fluid velocity $\vec{u}$, the magnetic induction $\vec{B}$ and the total energy density $e$. Additionally, $p$ is the thermal pressure, $\tens{I}$ is the unit tensor, $m_H$ is the mass of the hydrogen atom, and $\mu_0$ is the permeability of free space. The source terms on the right-hand side include  the vector of external force densities $\vec{f}$ and the radiative cooling term $\Lambda(T)$. The external forces are the gravitation by the individual stars and the radiation forces of the stars:
\begin{equation}
	\vec{f}
	=
	\rho
	\sum^n_{i=1} \left(
	- G M_{\star, i} \frac{\vec{r}_i}{r_i^3} 
	+ \vec{g}_{rad,i}^l
	+ \vec{g}_{rad,i}^e
	\right), 
\end{equation}
where the index $i$ relates to the individual stars. The vector $\vec{r}_i$ is the distance vector relative to star $i$ and $\vec{g}_{rad,i}^l$ and $\vec{g}_{rad,i}^e$ denote the radiative line acceleration by interaction with ions in the wind and acceleration by stellar radiation scattering off free electrons, respectively. Assuming that the acceleration is directed radially from the stars then leads to:
\begin{equation}
	\vec{g}_{rad,i}^e
	=
	g_{rad,i}^e \frac{\vec{r}_i}{r_i}
	\qquad
	\text{with}
	\qquad
	g_{rad,i}^e 
	= \frac{\sigma_e L_{\star,i}}{4\pi r_i^2 c}
\end{equation}
with $\sigma_e$ the mass-specific electron opacity due to Thomson-scattering. Taking into account the finite size of the stellar disk the term related to line acceleration $\vec{g}_{rad,i}^l$ can be expressed as:
\begin{equation}
	\vec{g}_{rad,i}^l
	=
	g_{rad,i}^l \frac{\vec{r}_i}{r_i}
	\qquad\text{with}\qquad
	g_{rad,i}^l = \frac{\sigma_e}{c} \frac{L_{\star,i}}{4\pi r_i^2}
	k t^{-\alpha} I_{FD}
\end{equation}
where $t$ is given as
\begin{equation}
	t = \left(\sigma_e \rho v_{th}\right)/\left|\frac{\partial u}{\partial r}\right|
\end{equation}
with $k$ and $\alpha$ the usual Castor-Abbot-Klein (CAK) parameters \citep[see][]{CastorAbbotKlein1975ApJ195_157}. The position dependent finite-disc correction factor $I_{FD}$ is determined from the radial velocity gradient and its projection onto the unit vector towards a point on the stellar surface $\vec{n}_i$:
\begin{equation}
	\label{EqFiniteDisc}
	I_{FD} = \frac{1}{1-\cos \theta_{\star,i}^2} \frac{1}{\pi}
	\int_0^{2\pi}
	\int_0^{\theta_{\star,i}}
	\left(
	\left(\vec{n}_i\cdot \nabla\left(\vec{n}_i\cdot\vec{u}\right)\right)
		/ \left|\frac{\partial u}{\partial r}\right|
		\right)^{\alpha} \cos \theta_i d \Omega_{i}.
\end{equation}
 This integral extents over the entire stellar disk as seen from position $\vec{r}$ and usually is solved numerically.
  At temperatures above $10^6$\,K line driving is set to zero since the plasma is highly ionised. For further details see, e.g., \citet{ReitbergerEtAl2014ApJ782_96,GayleyEtAl1997ApJ475_786}.

The temperature is initialised at a level of $10^4$\,K in the entire simulation domain. Cooling below this value is prevented numerically, emulating photo-ionization heating by the stellar radiation fields \citep[see][]{Pittard2009MNRAS396_1743}. Additionally we use the radiative cooling function by \citet{SchureEtAl2009AnA508_751}, which mainly becomes important for the shocked material within the WCR.

For the solution of the dynamical equations we use the \textsc{Cronos} code for numerical MHD \citep[see, e.g.,][]{KissmannEtAl2009JCP, KissmannEtAl2011ASPC444_36}. This code has been adapted for the use in CWB systems by including the CAK-force terms and the radiative cooling terms as detailed above \citep[see also][]{ReitbergerEtAl2014ApJ782_96}.

\subsection{The Near-star Resolution Problem}
As noted above, it is necessary to simulate the environment near the stellar surface sufficiently accurate to capture the impact of the stellar magnetic field on the acceleration of the wind. In \citet{udDoulaOwocki2002ApJ576_413} corresponding simulations are limited to a range from the stellar surface to six stellar radii to yield a detailed simulation of the wind acceleration, which is rather small compared to the size of a CWB system that can span thousands of solar radii. Therefore, we decided to use two distinct simulation setups - one for the acceleration region of the stellar wind for the individual stars and one for the CWB simulation. In the latter the detailed simulations for the individual stellar winds were used as input, which was kept constant throughout the simulation -- just like the constant beta law prescription in the hydrodynamic simulations. Both numerical setups will be described below.

\subsubsection{Near-star Magnetized Wind Models}
For the simulation of stellar winds near the surface of the individual stars the effect of the respective other star is ignored, later to be considered in the full CWB simulations. The purpose of the single-star simulations is to obtain the structure of the stellar winds near the stellar surface including the back reaction of the magnetic field on the winds. The simulation setups are similar to the ones detailed in \citet{udDoulaOwocki2002ApJ576_413} with regard to the computational grid, the boundary conditions, and our use of an isothermal stellar-wind outflow. The only relevant difference is our use of a somewhat more general finite disc correction factor detailed below.


Computation of the near-star wind models is also a two-step process. In a first step, the CAK parameters for these simulations are obtained by corresponding one-dimensional spherically-symmetric hydrodynamical models. These models use the same radial grid and the same boundary conditions for the hydrodynamic variables as in the two-dimensional magnetized simulations, with the exception that the grid is further extended to some 1000 $R_{\sun}$ to assure the correct estimate for the terminal velocity. During these hydrodynamical simulations $k$ and $\alpha$ are adapted alternatingly until a wind solution is found that gives the desired values for the mass-loss rate and the terminal velocity. Here, $k$ and $\alpha$ are adapted up to an accuracy of $10^{-5}$. 

In the second step these values for the CAK parameters are used in two-dimensional simulations including the stellar magnetic field. In this initial study without stellar rotation we still assume strict axial symmetry. Since we use the CAK parameters from hydrodynamical simulations it is not unlikely -- and for a sufficiently strong magnetic field indeed expected -- that the structure of the stellar winds will become different than in the hydrodynamic case.

For these two-dimensional simulations, the numerical mesh spans from the stellar surface to a maximum distance of about $6 R_{\star}$. We use a non-linear grid in the radial direction, which increases outwards by 2\% per cell. The first cell starting at the stellar surface is set to have a size of $10^{-3}R_{\star}$, where we use a total of 256 cells in the radial direction. In the $\theta$ direction we use 180 equidistant cells. This does not yield such a high resolution near the equator as used by \citet{udDoulaOwocki2002ApJ576_413}. However, we consider stellar magnetic fields  with confinement parameters $\eta < 1$ that do not lead to such a strong compression of the flow structure near the equator as for $\eta>1$ (in SI units the magnetic confinement parameter is given as:
\begin{equation}
	\eta = \pi \frac{B_{p}^2 R_{\star}^2}{\dot M v_{\infty}}
\end{equation}
where $B_{p}$ is the strength of the surface magnetic field at the pole). Apart from that, the focus of this study is not on the acceleration of the stellar wind near the stars but rather on the impact on the large-scale flow structure including the corresponding magnetic field at larger distances from the star.


For the boundary conditions we use an extrapolation at the outer radial boundaries of the numerical domain, where the flow is supersonic. On the $z$-axis of the spherical coordinate system we use adapted axis-boundary conditions \citep[see][]{Ziegler2011JCP230_1035}. At the stellar surface, the density is kept fixed, while the velocity is allowed to vary in the first ghost cell only. Here, linear extrapolation of the mass-flux into the ghost cell is used. For the magnetic field we prescribe fixed vector-potential components within the star that represent a dipole field aligned with the $z$-axis. Additionally, the temperature is set to $10^4$\,K throughout the entire numerical domain.

The two-dimensional magnetized simulations are initialized with a linearly increasing velocity where the density is given locally by the desired mass-loss rate. Additionally, a dipole field is initialized within the entire numerical domain. From these initial conditions the wind is allowed to evolve until it converges to a steady state. In contrast to the one-dimensional spherically-symmetric simulations, the finite disk correction factor needs to be computed numerically in the two-dimensional magnetized simulations. This is because of the possibly non-zero velocity component in the $\theta$ direction and because a possible variation of both velocity components in the $\theta$ direction needs to be taken into account. For this we use:
\begin{align}
	\vec{n}\cdot \nabla \left(\vec{n}\cdot\vec{v}\right)
	=&
	\left(\frac{\partial v_r}{\partial r}\right) \cos \theta^2
	+
	\left(\frac{1}{r} \frac{\partial v_{\theta}}{\partial \theta} + \frac{v_r}{r}\right) \sin \theta^2 \cos \phi^2
	\nonumber\\
	&+
	\left(
	\frac{v_r}{r} + \frac{v_{\theta} \cot \theta_0}{r}
	\right)
	\sin \theta^2 \sin \phi^2
	+
	\left(
	\frac{\partial v_{\theta}}{\partial r}
	- \frac{v_{\theta}}{r} + \frac{1}{r}\frac{\partial v_r}{\partial \theta}
	\right)
	\cos \theta \sin \theta \cos \phi
\end{align}
within Eq. \eqref{EqFiniteDisc} as taken from Eq. (41) in \citet{CranmerOwocki1995ApJ440_308}.

The simulation is stopped when the mass-loss rate is identical at all radii, indicating that a steady-state solution has been reached. The mass-loss rate is computed from the numerical mass flux through the outer radial boundary of each cell. The results of these two-dimensional simulations are stored as reference solutions for the CWB simulations.

\subsubsection{Numerical Setup of Colliding-wind Binary Simulations}
The CWB simulations are done on a three-dimensional Cartesian mesh with a resolution of 512 cells in each spatial dimension. The stars are located on the $x$-axis with the same distance from the center of the computational domain. The size of the numerical domain is varied with the separation $d$ of the stars, where models with $d=720$\,$R_{\Sun}$, $d=1440$\,$R_{\Sun}$, and $d=2880$\,$R_{\Sun}$ are investigated for which we use an extent of $-500\dots 500$, $-1000\dots 1000$, and $-2000\dots 2000$\,$R_{\Sun}$ of the numerical domain, respectively. On all boundaries simple extrapolation is used since the flow is supersonic at the boundaries even inside the WCR.

The simulations are initialized by mapping the reference solutions from the near-star simulation models onto the coarser numerical mesh of the large-scale simulations. The hypothetical position of the WCR is computed from the parameters of the undisturbed hydrodynamical winds. By this the domain is subdivided into two-parts \citep[see also][]{ReitbergerEtAl2014ApJ782_96}. In each part the wind is initialized from the values of the respective near-star simulation. Since the scale of the CWB simulations is much larger than the one for the near-star simulations, cells outside the range of the near-star solution are filled with the values at the largest radius of the near-star solution. The wind is initialized with a temperature of $10^4$\,K.
 
After initialization the stellar winds are allowed to evolve freely. The solution from the near-star simulations is fixed from each star's surface up to 30 $R_{\sun}$ above the surface. While this prevents an effect of the other star's radiation on the wind near a star's surface, it is used as a compromise to conserve the fine-structure for the wind and the magnetic field in the vicinity of the stars. In the CWB simulations the finite disc correction factor is computed numerically.


\section{Results}
\label{SecResults}
\subsection{Investigated Models}
\label{SecSimSingle}
In this study the binary system composition (B + WR star) as in \citet{ReitbergerEtAl2014ApJ782_96,ReitbergerEtAl2014ApJ789_87} is considered with stellar parameters given in Table \ref{TabStars}, where orbital motion is neglected. In this study we consider nine different binary systems with the relevant system parameters given in Table \ref{TabSetups}. We consider three systems having different stellar separations $d$ with $d=720$\,$R_{\sun}, 1440$\,$R_{\sun}, \text{and } 2880$\,$R_{\sun}$ with dipole axes of the stars along the $z$-direction (models A1, A2, and A3). Additionally, three models with $d=1440$\,$R_{\sun}$ and different inclinations of the magnetic dipole axes ($\delta_{\text{B}}$ and $\delta_{\text{WR}}$, respectively) are investigated (models B1, B2, and B3). 


\begin{table}
\begin{tabular}{lcccccccc}
\hline\hline
Star & $M_{\star}$ & $R_{\star}$ & $T_{eff,\star}$ & $L_{\star}$ & 
$\dot M$ & $v_{\infty}$ & $\alpha$ & $k$
\\
& ($M_{\sun}$) & ($R_{\sun}$) & (K) & ($L_{\sun}$) & 
($M_{\sun} \text{yr}^{-1}$) & (m\,s$^{-1}$) &  &  
\\
\hline
B & 30 & 20 & 23000 & $10^5$ & $10^{-6}$ & $4\cdot 10^6$ & 0.68782 & 0.269452\\
WR & 30 & 10 & 40000 & $2.3\cdot 10^5$ & $10^{-5}$ & $4\cdot 10^6$ & 0.62351 & 0.881606 \\
\hline
\end{tabular}
\caption{\label{TabStars} Stellar and stellar-wind parameters for both stars.}
\end{table}

\begin{table}
\begin{tabular}{lccccccccc}
\hline\hline
Model: & A1 & A2 & A3 & B1 & B2 & B3 & C1 & C2 & C3\\
\hline
$d[R_{\sun}]$ &720 & 1440 & 2880 & 1440 & 1440 & 1440 & 1440 & 1440 & 1440
\\
$\delta_{\text{B}}[\text{degrees}]$ & 0 & 0 & 0 & 5 & 0 & 30 & 0 & 0 & 0
\\
$\delta_{\text{WR}}[\text{degrees}]$ & 0 & 0 & 0 & 0 & 30 & -40 & 0 & 0 & 0
\\
$B_{\text{polar}} [T]$ & $10^{-2}$ & $10^{-2}$ & $10^{-2} $
& $10^{-2}$ & $10^{-2}$ & $10^{-2}$
& $5\cdot 10^{-3}$ & $1.5\cdot 10^{-2}$ & $2\cdot 10^{-2}$\\
\hline
\hline
\end{tabular}
\caption{\label{TabSetups} Setup of colliding-wind binary systems.}
\end{table}

\begin{figure}[t]
	\setlength{\unitlength}{0.00033\textwidth}
	\begin{picture}(900,1000)(-100,0)
	\put(-70,420){\tiny\rotatebox{90}{$R$ [$R_{\sun}$]}}
	\includegraphics[height=1000\unitlength]{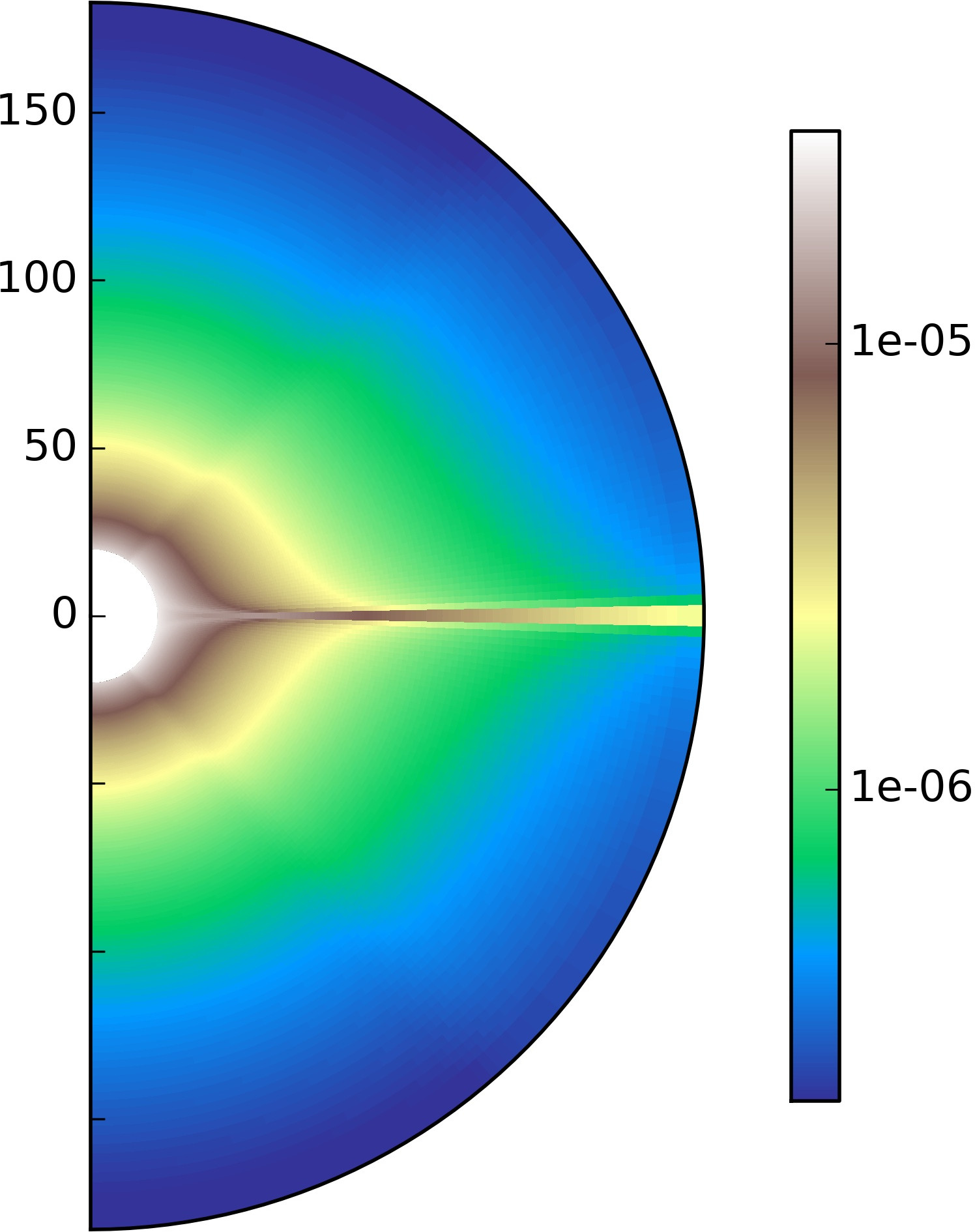}
	\put(-300,930){\tiny $\rho |\vec{u}|$ [kg\,m$^{-2}$ s$^{-1}$]}
	\end{picture}
	\hfill
	\begin{picture}(810,1000)(0,0)
	\includegraphics[height=1000\unitlength]{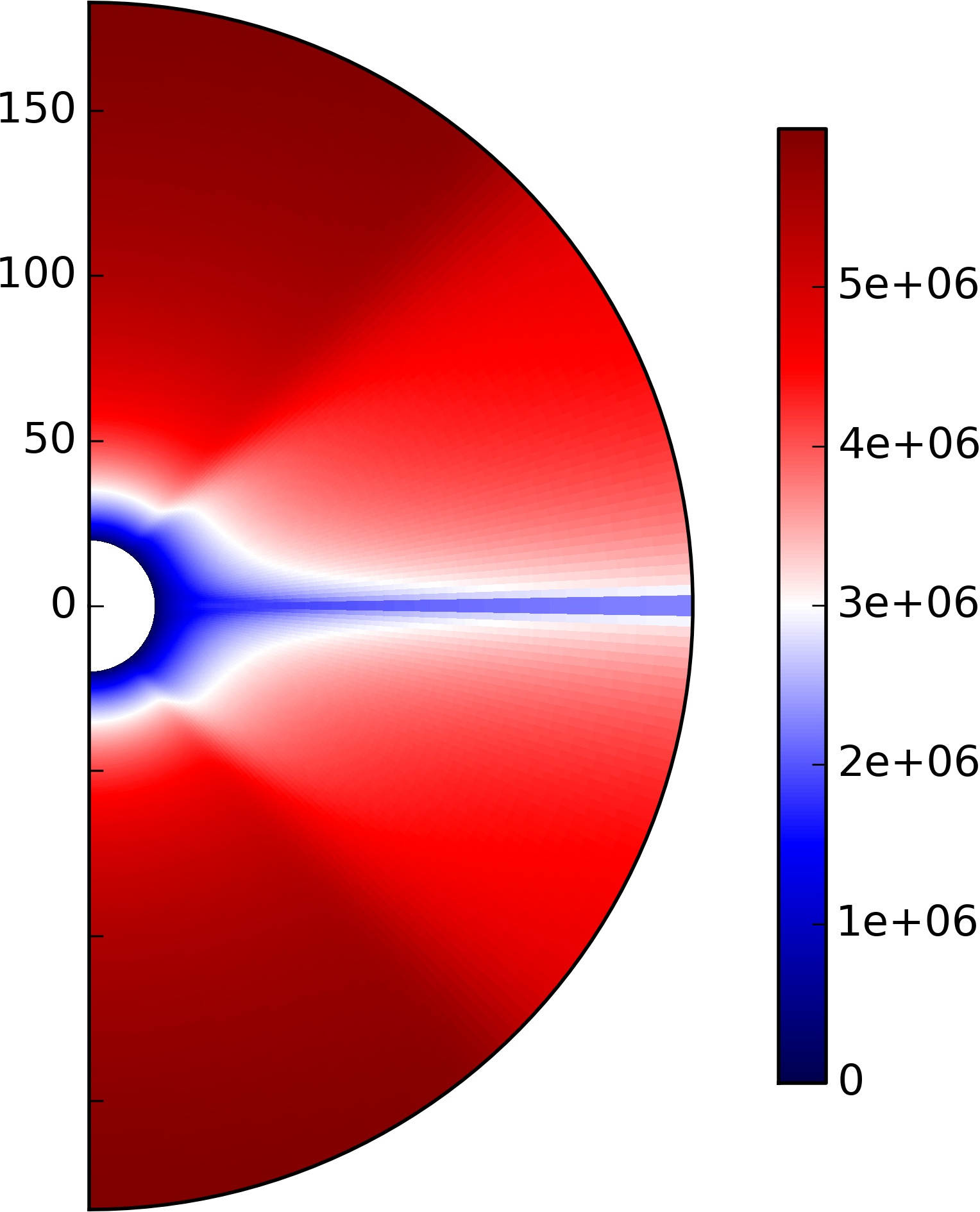}
	\put(-300,930){\tiny $|\vec{u}|$ [m\,s$^{-1}$]}
	\end{picture}
	\hfill
	\begin{picture}(820,1000)(0,0)
	\includegraphics[height=1000\unitlength]{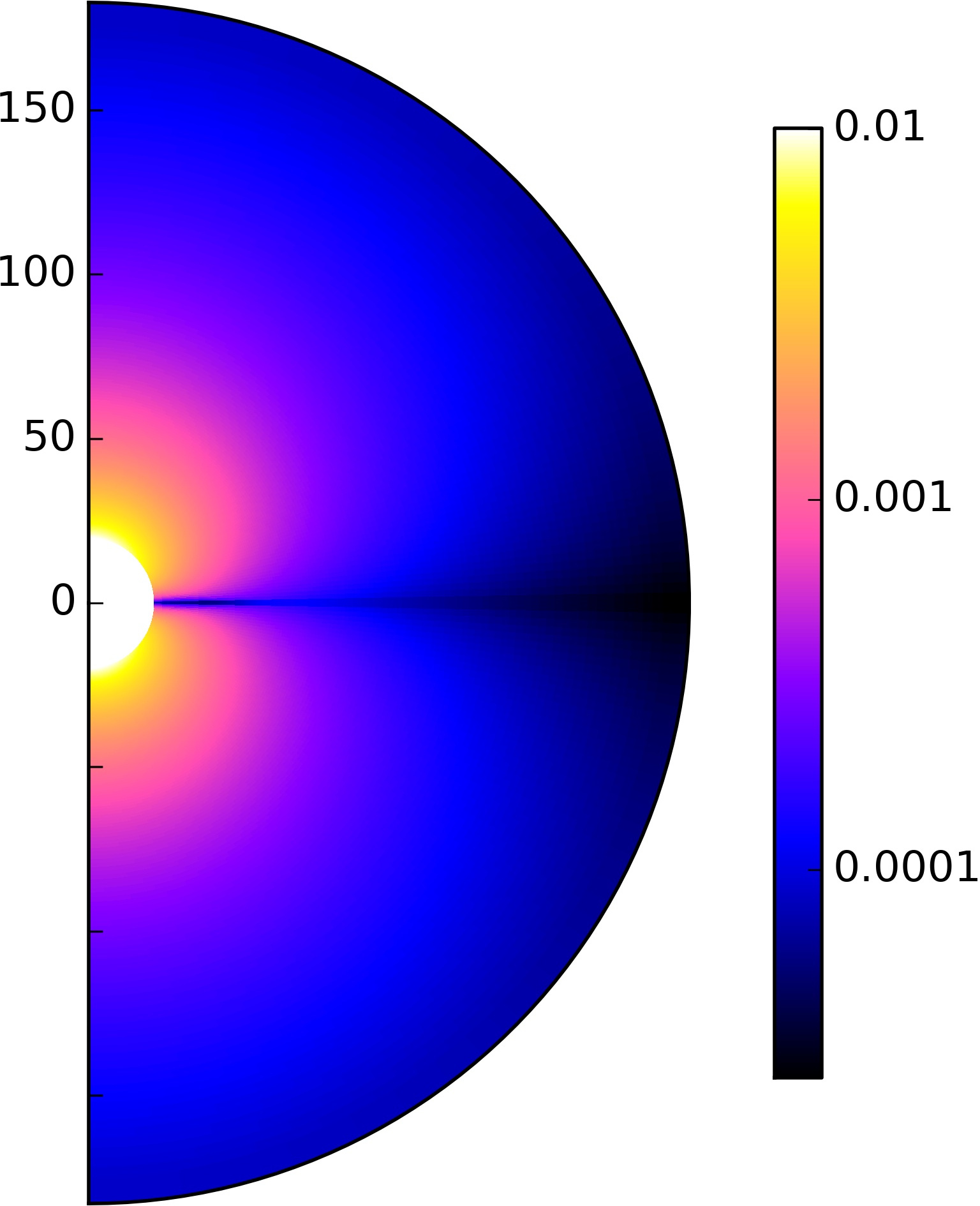}
	\put(-240,930){\tiny $|\vec{B}|$ [T]}
	\end{picture}
	\caption{\label{FigRefModelBStar}Results for the B-star reference model with a polar field strength of $0.01$\,T. The dipole field is aligned with the vertical axis. Results are shown for the momentum density (left), the absolute value of the wind velocity (middle) and the absolute value of the magnetic induction (right).}
\end{figure}

In all these simulations we used a polar magnetic field strength of $10^{-2}$\,T for both stars. Together with the wind momentum (see Table \ref{TabStars}) this leads to a magnetic confinement parameter of $\eta = 0.192$ for the B star and $\eta = 0.0048$ for the WR star, respectively. Thus, a significant influence of the magnetic field on the stellar wind outflow can only be expected for the B star. This is indeed found from the near-star simulation results. Fig. \ref{FigRefModelBStar} shows converged results for the B-star model. The influence of the magnetic field is apparent in the latitudinal dependence both of velocity and momentum density. Here, we find qualitatively similar behavior as was observed by \citet{udDoulaOwocki2002ApJ576_413} for their $\eta=0.1$ case with a somewhat stronger deflection of the flow towards the magnetic equator in our simulations. 
This deflection leads to a peak of the radial mass flux density around the magnetic equator of the star. Also, in our simulations we find an increase of the wind speed above the poles of the stars (see middle panel in Fig. \ref{FigRefModelBStar} in contrast to the terminal velocity of $4\cdot 10^6$ m/s in the hydrodynamic case). This, together with the peak of the radial mass flux at the equator has a considerable influence on the structure of the WCR as is discussed below.

Additionally, we investigate models with polar magnetic field strengths of $5\cdot10^{-3}$\,T, $1.5\cdot10^{-2}$\,T, and $2\cdot10^{-2}$\,T (models C1, C2, and C3). These lead to magnetic confinement parameters of $\eta = 0.048, 0.432, \text{and } 0.768$ for the B star and $\eta = 0.0012, 0.011, \text{and }0.0192$  for the WR star, respectively. These models use a stellar separation of 1440\,$R_{\sun}$ and an orientation of the dipole axes along the $z$-direction. As in the other cases, only the B star's wind is significantly affected by the presence of the magnetic field, where in the model with a polar magnetic field strength of $5\cdot10^{-3}$\,T this influence is only minor.

\subsection{Structure of magnetised CWBs}
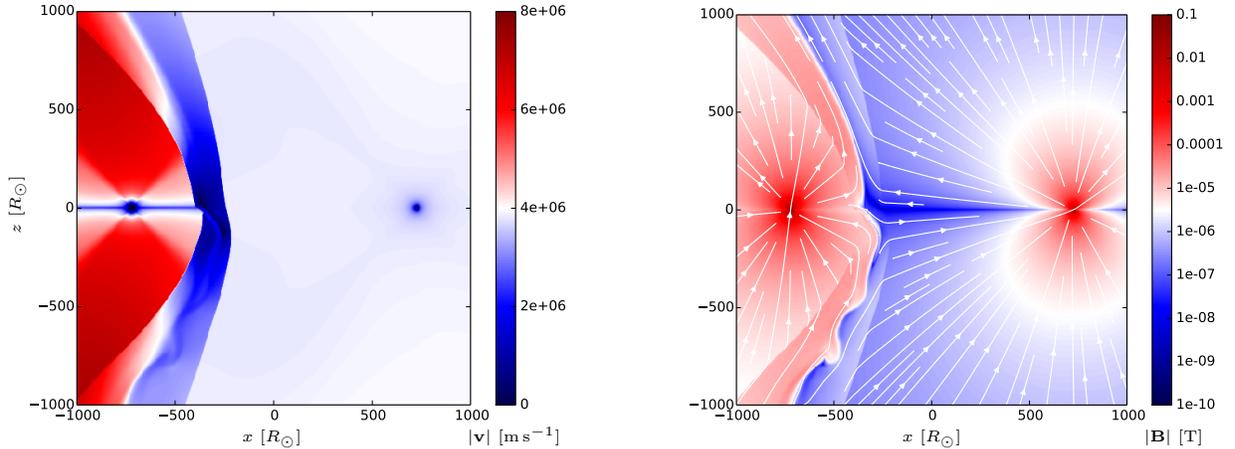
\begin{figure}[ht]
	\setlength{\unitlength}{0.00034\textwidth}
	\begin{picture}(1424,1100)(-100,-100)
	\put(-50,450){\tiny\rotatebox{90}{$z$ [$R_{\sun}$]}}
	\put(510,-50){\tiny$x$ [$R_{\sun}$]}
	\put(1050,-50){\tiny$|\vec{v}|$ [m\,s$^{-1}$]}
	\includegraphics[height=1000\unitlength]{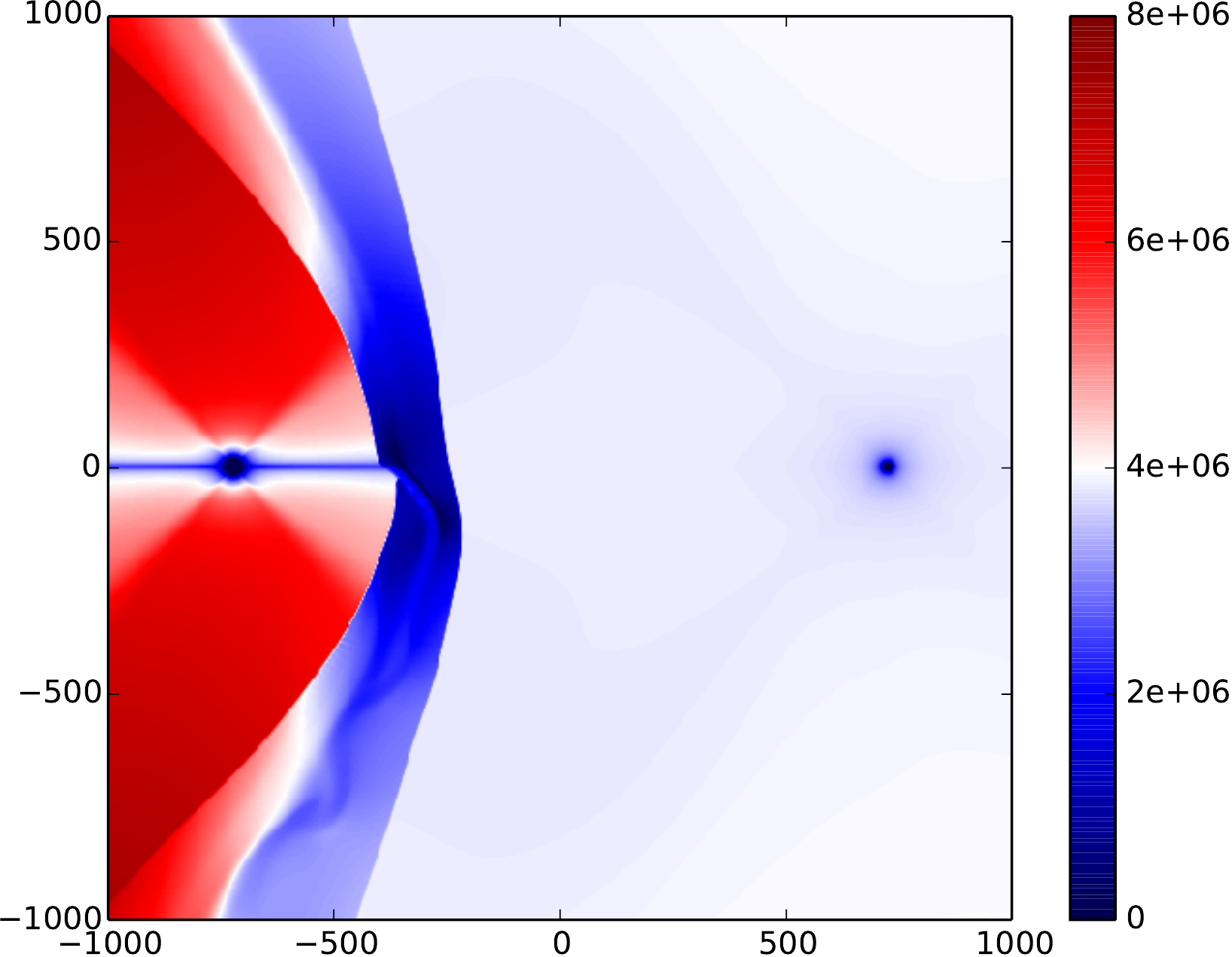}
	\end{picture}
	\hfill
	\begin{picture}(1296,1100)(0,-100)
	\put(510,-50){\tiny$x$ [$R_{\sun}$]}
	\put(1090,-50){\tiny$|\vec{B}|$ [T]}
	\includegraphics[height=992\unitlength]{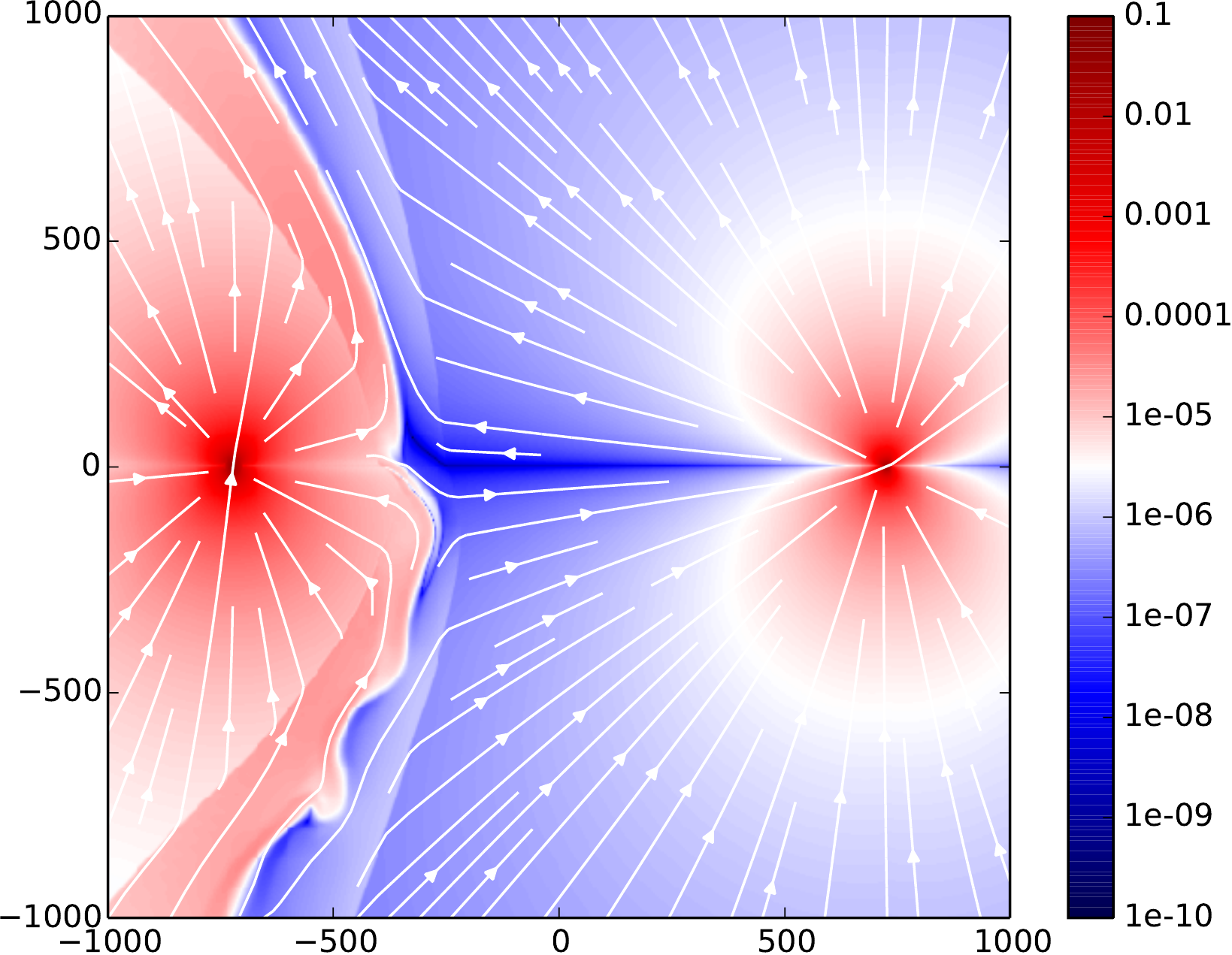}
	\end{picture}
	\caption{\label{FigCWBModelB}Results for a setup of the magnetized CWB with a stellar separation of 1440\,$R_{\sun}$ using a polar magnetic field strength of $10^{-2}$\,T for both stars. Simulations results are shown for the velocity (left) and the magnetic induction (right) in the $x-z$ plane. In each case the absolute values are shown. For the magnetic field we also show the vector direction.}
\end{figure}

As an example, results for model A2 with a stellar separation of 1440\,$R_{\sun}$ are shown in Fig. \ref{FigCWBModelB}. Here, results are shown in a slice through the three-dimensional computational domain, spanned by the magnetic dipole axes of the stars and the line of centers between both stars. Expectedly, the structure of the WCR becomes more complex in the presence of the magnetic field: in the bottom half of the image the contact discontinuity is affected by turbulence, while in the upper half the flow in the WCR is laminar. This asymmetry between upper and lower half is remarkable since the initial simulation setup is symmetric.



In the discussion of the time evolution of the simulations, we show that this asymmetry relates to an instability in the equatorial region, i.e. to a region around the magnetic current sheet of the B star. It is related to the higher mass-loss rate in the equatorial region of the B star, which is an effect of the presence of the magnetic field as already discussed in \citet{udDoulaOwocki2002ApJ576_413} (see also Fig. \ref{FigRefModelBStar} on the left). There the current sheet has been found to be prone to oscillations even in single-star simulations, but this effect only arises for stronger magnetic fields than considered here. In the CWB systems, however, the interaction with the other star's wind seems to boost this instability leading to the asymmetry visible in Fig. \ref{FigCWBModelB}. 

In the following discussion we show that the presence of the magnetic field leads to higher turbulence levels in the WCR than in the pure hydrodynamic case: for the models investigated here, the WCR shows a laminar flow in the hydrodynamic case, while it becomes at least partly turbulent for the highly magnetised winds. We found that the growth rate of the related instability increases for the case of the magnetised stellar winds, thus, explaining the occurrence of turbulence. For simulations showing a laminar flow in the WCR the growth rates of the instability driving the turbulence is too low to lead to a turbulent WCR. Since the growth rate for the Kelvin-Helmholtz instability rises with decreasing wavelength, e.g., also the upper half of the WCR shown in Fig. \ref{FigCWBModelB} will become turbulent in higher resolution simulations.

\subsection{The Emergence of Asymmetry}
\label{SecTimeDep}
\begin{figure}[h!]
\begin{center}
	\setlength{\unitlength}{0.00026\textwidth}
	\begin{picture}(1202,1100)(-100,-100)
	\put(-50,400){\tiny\rotatebox{90}{$z$ [$R_{\sun}$]}}
	\put(500,-50){\tiny$x$ [$R_{\sun}$]}
	\includegraphics[height=1000\unitlength]{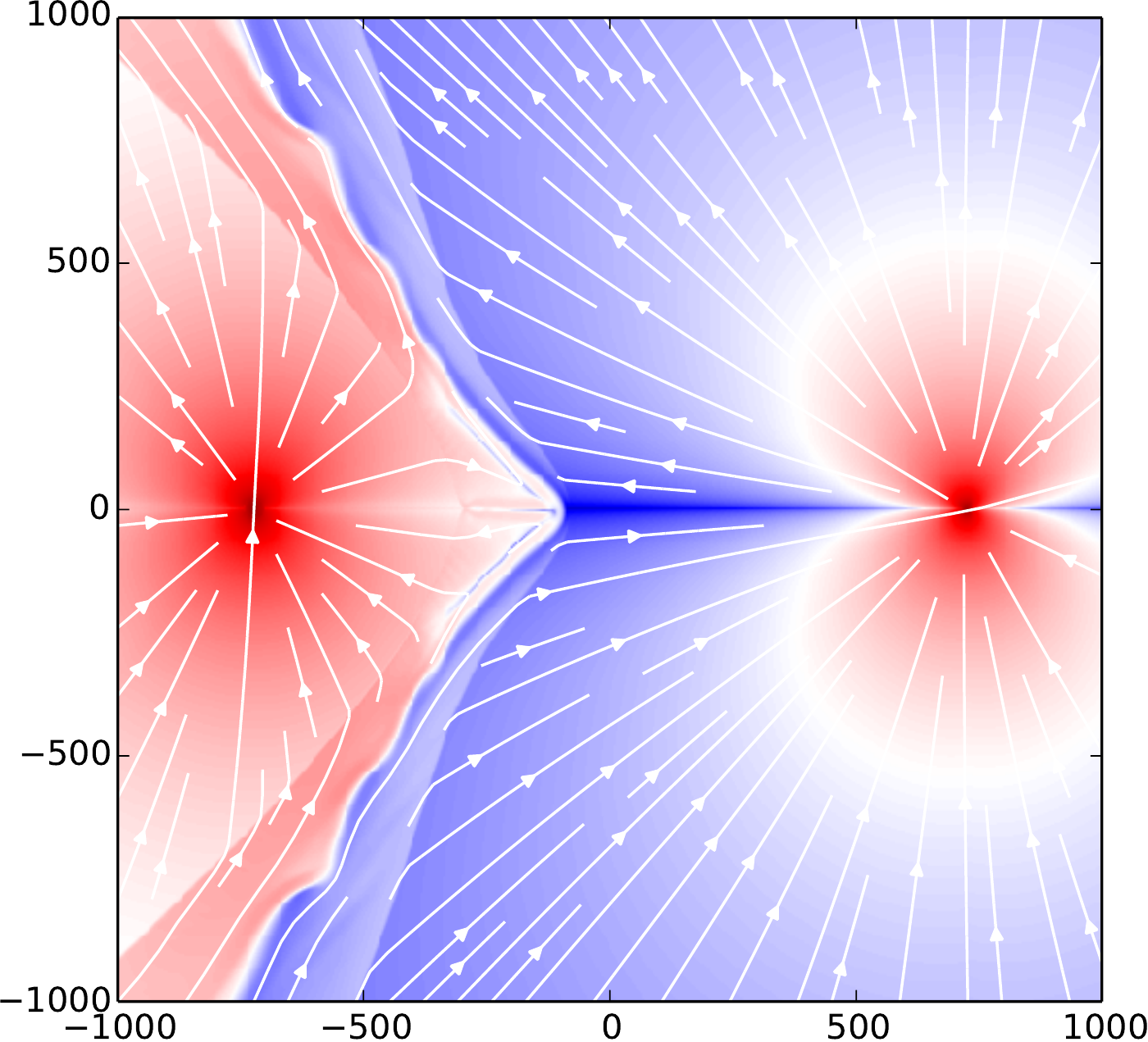}
	
	\end{picture}
	\hfill
	\begin{picture}(1102,1100)(0,-100)
	\put(500,-50){\tiny$x$ [$R_{\sun}$]}
		\includegraphics[height=1000\unitlength]{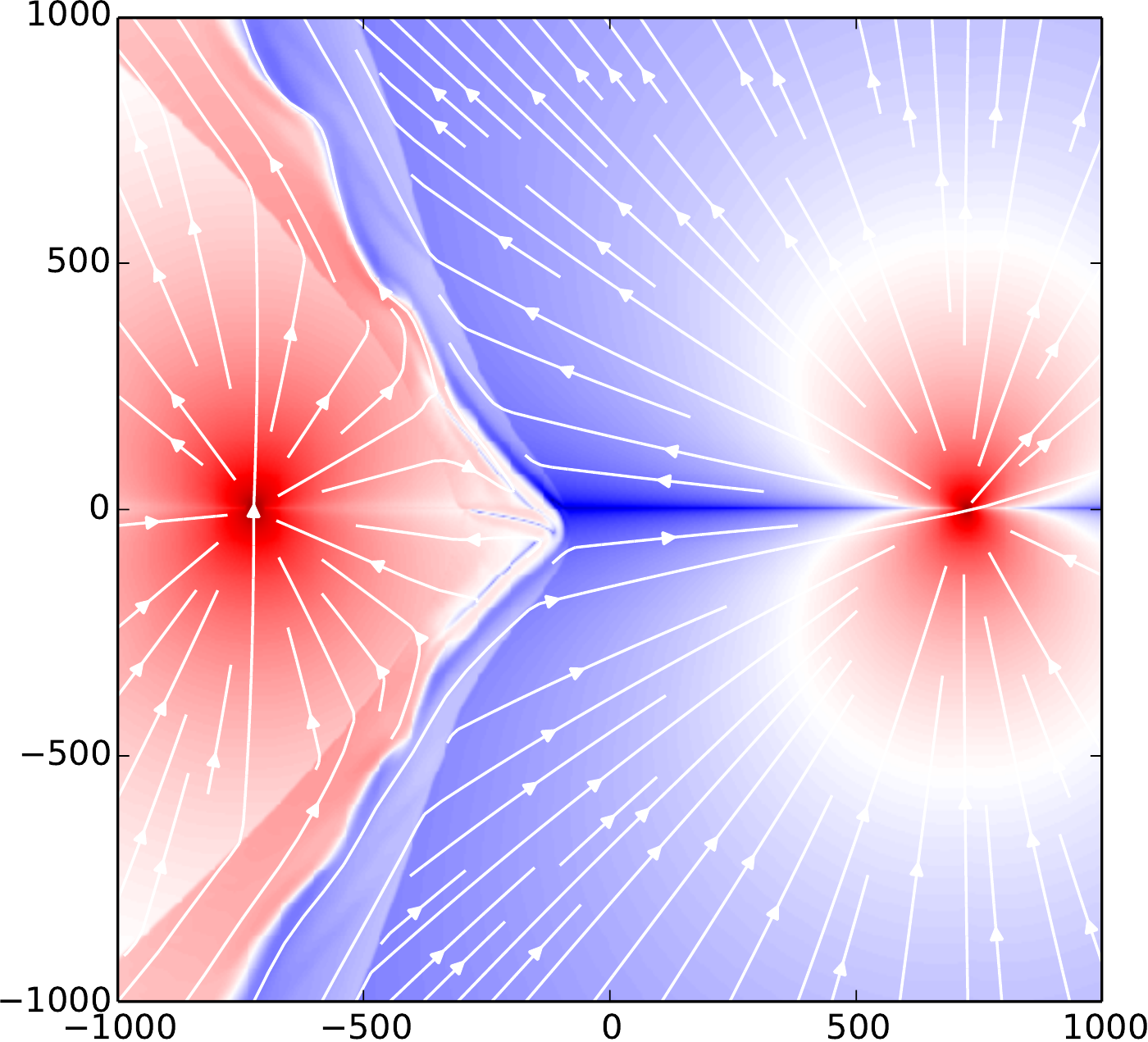}
	\end{picture}
	\hfill
	\begin{picture}(1291,1100)(0,-100)
	\put(500,-50){\tiny$x$ [$R_{\sun}$]}	
	\put(1090,-50){\tiny$|\vec{B}|$ [T]}
	\includegraphics[height=1000\unitlength]{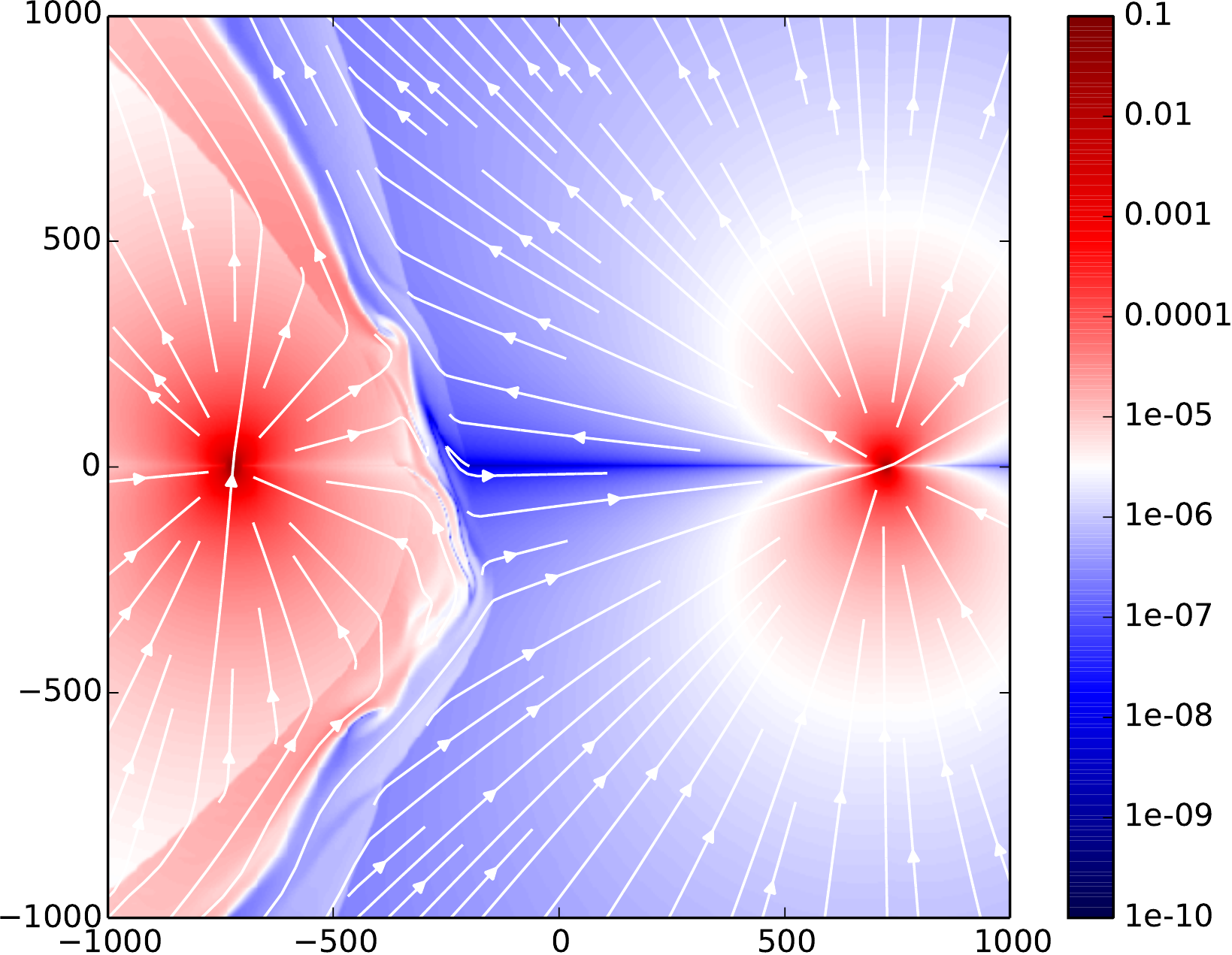}
	\end{picture}
	
	\begin{picture}(1202,1100)(-100,-100)
	\put(-50,400){\tiny\rotatebox{90}{$z$ [$R_{\sun}$]}}
	\put(500,-50){\tiny$x$ [$R_{\sun}$]}
	\includegraphics[height=1000\unitlength]{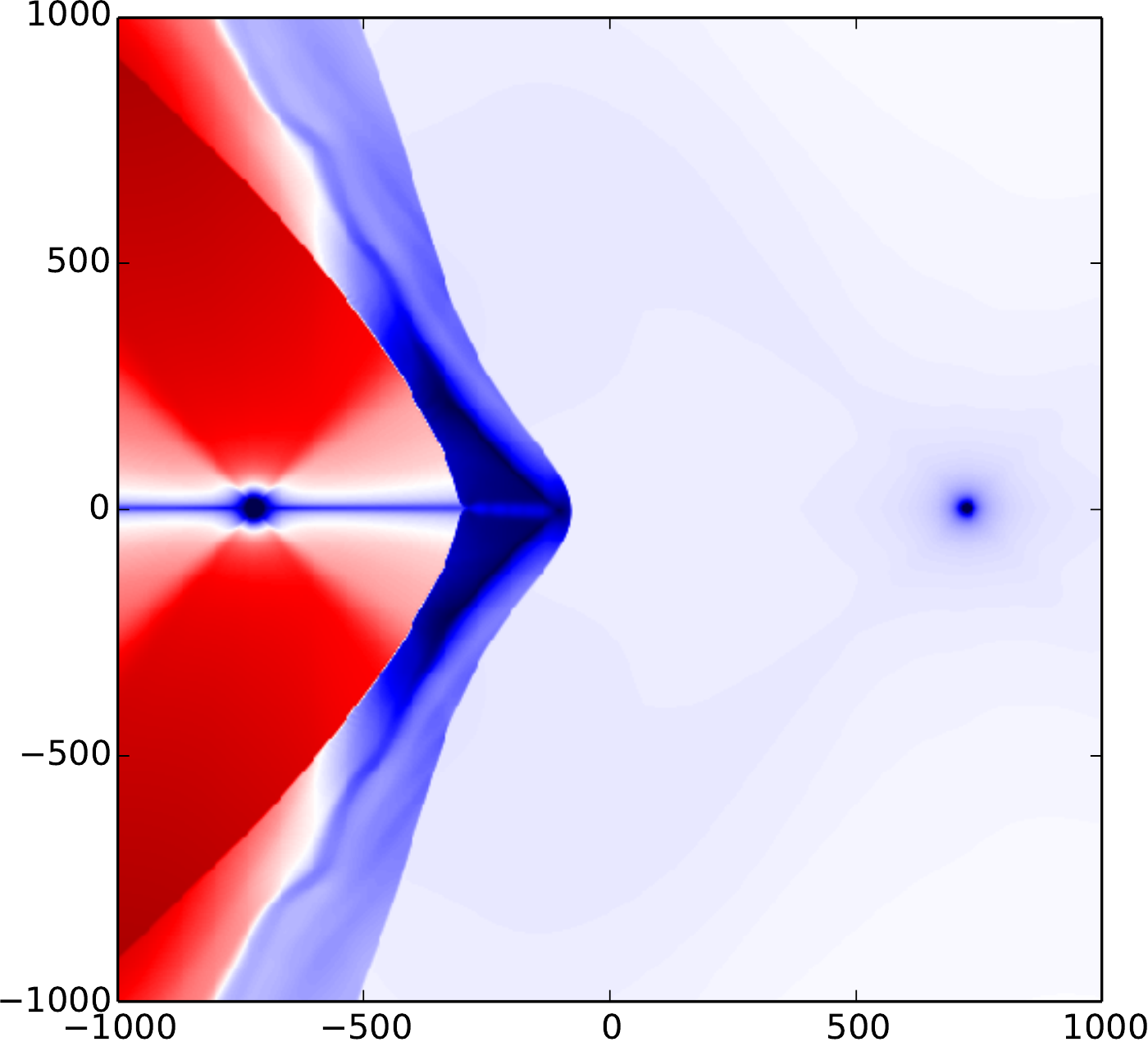}
	\end{picture}
	\hfill
	\begin{picture}(1102,1100)(0,-100)
	\put(500,-50){\tiny$x$ [$R_{\sun}$]}
	\includegraphics[height=1000\unitlength]{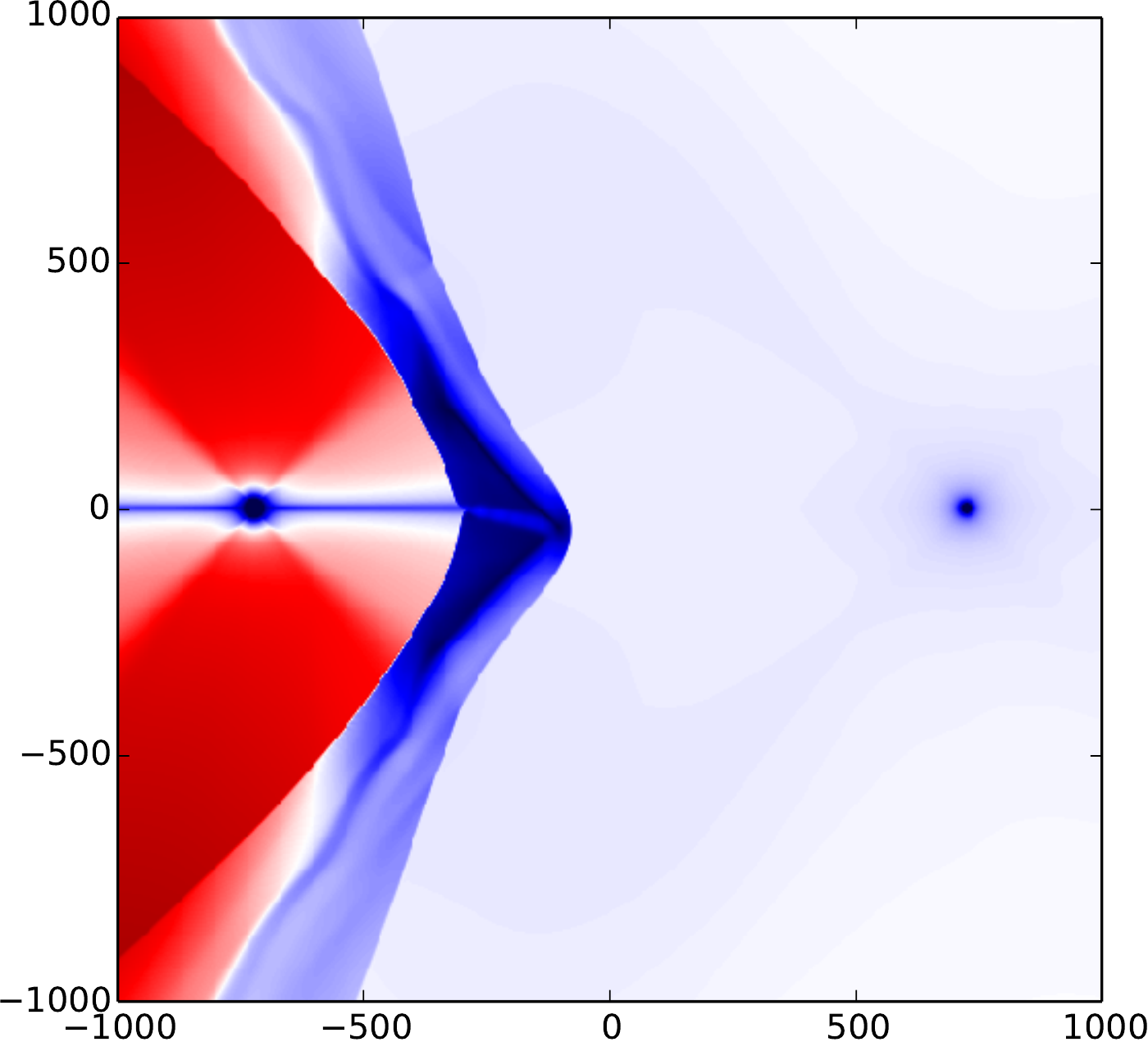}
	\end{picture}
	\hfill
	\begin{picture}(1291,1100)(0,-100)
	\put(500,-50){\tiny$x$ [$R_{\sun}$]}
	\put(1050,-50){\tiny$|\vec{v}|$ [m\,s$^{-1}$]}
	\includegraphics[height=1000\unitlength]{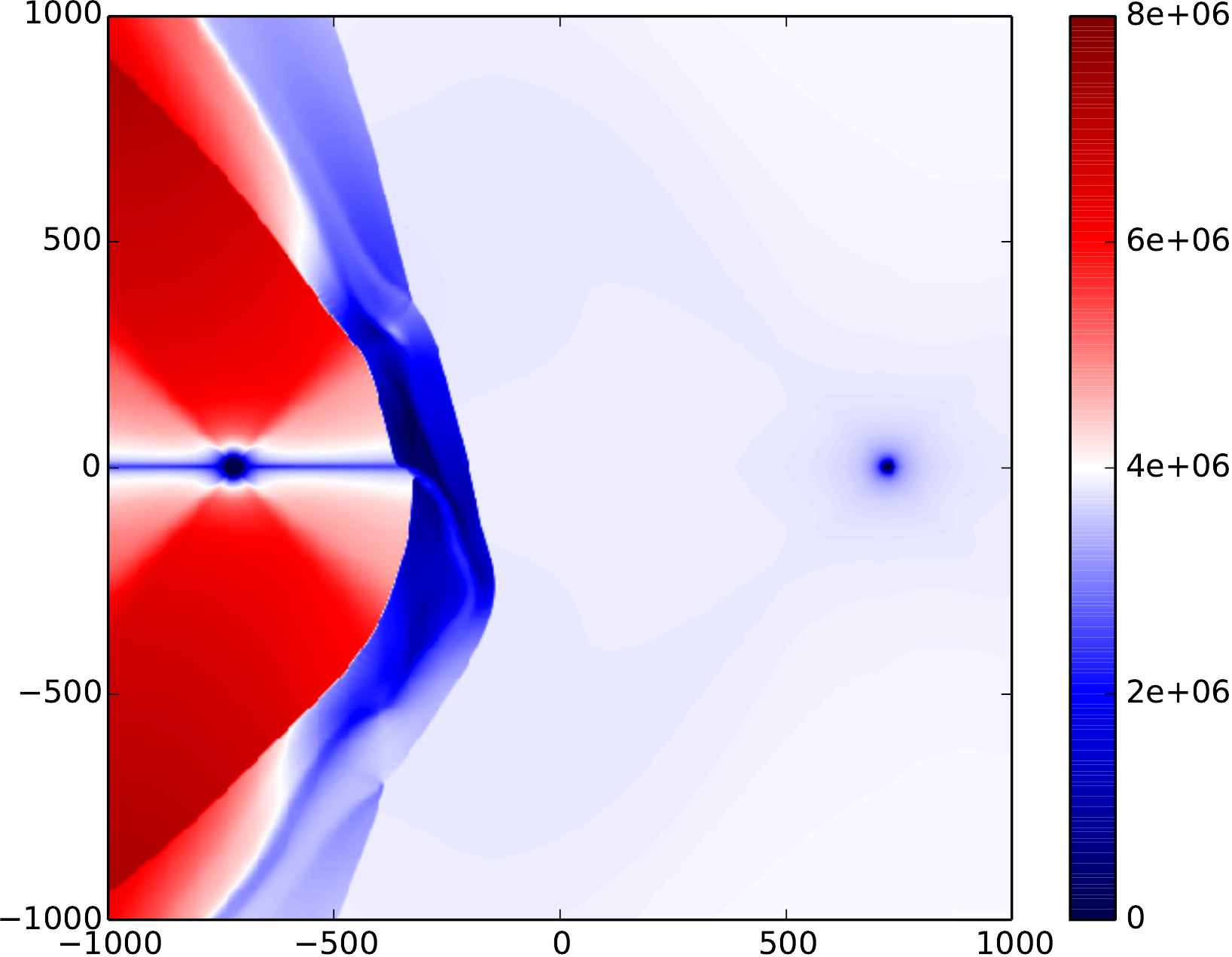}
	\end{picture}
	\end{center}
		\caption{\label{FigModelBvartime}Results for the magnetized CWB simulation for setup A2 for the magnetic induction (top) and the velocity (bottom) in the $xz$-plane at three different time steps. These correspond to a physical system time of 13.7, 17.2 and 20.6 days after the start of the simulation. The ``converged'' result in Figure \ref{FigCWBModelB} shows a time of 27.5 days after initiation. Both stars have a polar magnetic field strength of 0.01\,T.
In each case the absolute value is shown. For the magnetic induction we also show the vector direction. The entire time sequence is available as an animation in the electronic edition of the \emph{Astrophysical Journal}.}
\end{figure}
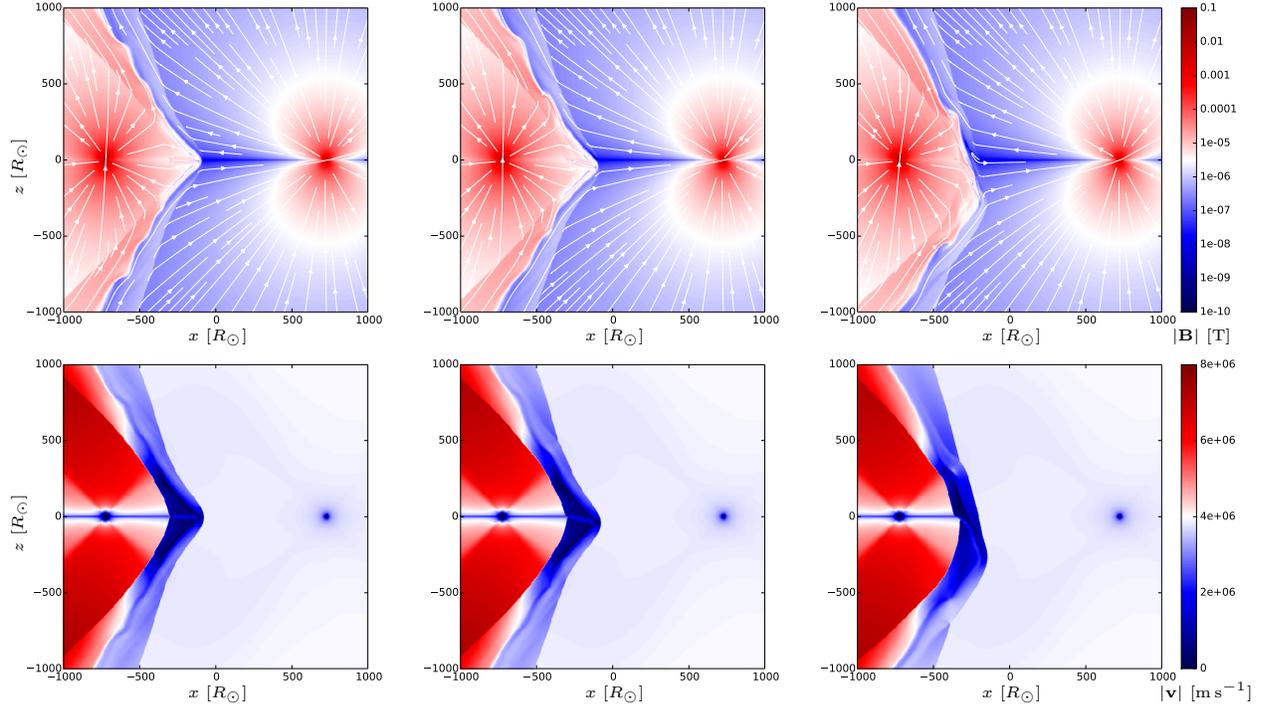

One of the most intriguing features of the results discussed above is their apparent asymmetry along the $z$-axis. It is astonishing that a setup in which all initial properties and acting forces above the $xy$-plane are an exact mirror image of the ones below should develop such pronounced asymmetry. 

To address the question of how these asymmetries emerge, Fig. \ref{FigModelBvartime} shows a time sequence for setup A2 leading up to the results shown in Fig. \ref{FigCWBModelB}. The magnetic induction and the velocity field are displayed in the $xz$-plane for three different time steps before convergence is reached. In the first column of Fig. \ref{FigModelBvartime} the WCR is still symmetrical. Apparently, the higher mass-loss rate around the magnetic equator of the B star causes a pointed \emph{nose}-like feature at the apex. At this state the simulation remains stable for a fair amount of time as if feigning to have already converged. 

The center column in Fig. \ref{FigModelBvartime} shows a state where symmetry begins to fade. Progressively the nose slides in negative $z$-direction and finally collapses into the remaining WCR, leaving behind the apparent asymmetry (Fig. \ref{FigModelBvartime}, right column). The direction into which the nose begins to slide is found to be of stochastic nature. 

The instability causing the sliding of the nose appears to be of Kelvin Helmholtz (KH) type \citep[see also][for a discussion on the occurrence of the KH instability in colliding-wind binary systems]{LambertsEtAl2011MNRAS418_2618}. The shear velocity becomes particularly high due to the presence of the nose structure. On the one hand, the fluid on the B star's side of the contact discontinuity of the WCR comes to a near stop for a rather broad region around the magnetic current sheet. On the other hand the central peak of the WCR on the WR star's side leads to a higher velocity of the WR star's wind along the contact discontinuity. 

To ensure that the KH instability can explain the observed fluctuations we ensured its non-suppression in presence of magnetic fields. For the setup  depicted in Fig. \ref{FigModelBvartime} at system time of 13.7 days we computed the ratio:
\begin{equation}
 \xi = \frac{\alpha_l \alpha_r(v_l - v_r)^2}{B^2/\bar \rho }
 \qquad\text{with}\qquad
 \alpha_i = \frac{\rho_i}{\rho_l + \rho_r}
\end{equation}
at the contact discontinuity where the presence of the magnetic field can suppress the KH instability for $\xi<1$ \citep[see][]{Chandrasekhar1961Book}. Here indexes $r$ and $l$ indicate density $\rho$ and velocity $v$ on the right and left of the contact discontinuity, respectively. Up to  $\sim$400 $R_{\sun}$ above the line of centers between the stars, however, $\xi$ is several orders of magnitude larger than unity.

Correspondingly, we did a quantitative analysis of the growth rate for the KH instability. The growth rate is particularly large around the region of the nose feature -- up to about $250 R_{\sun}$ above and below the line of centers between both stars. There, the growth timescale for fluctuations of a wavelength $\lambda = 100 R_{\sun}$ can be nearly as small as a tenth of a day. With the timescale for a fluid element to propagate through this strongly unstable region being longer than a day the KH instability works sufficiently rapid to allow for flipping of the nose structure and the occurrence of fluctuations in the outer parts of the WCR. This also shows that the presence of the magnetic field leads to a more unstable contact discontinuity as compared to the hydrodynamic case. Once flipping of the nose structure happend, the growth rate of the KH instability decreases approximately by a factor of four, but is still largest in the central part of the WCR.




\subsection{Models for Different Separations of the Stars}
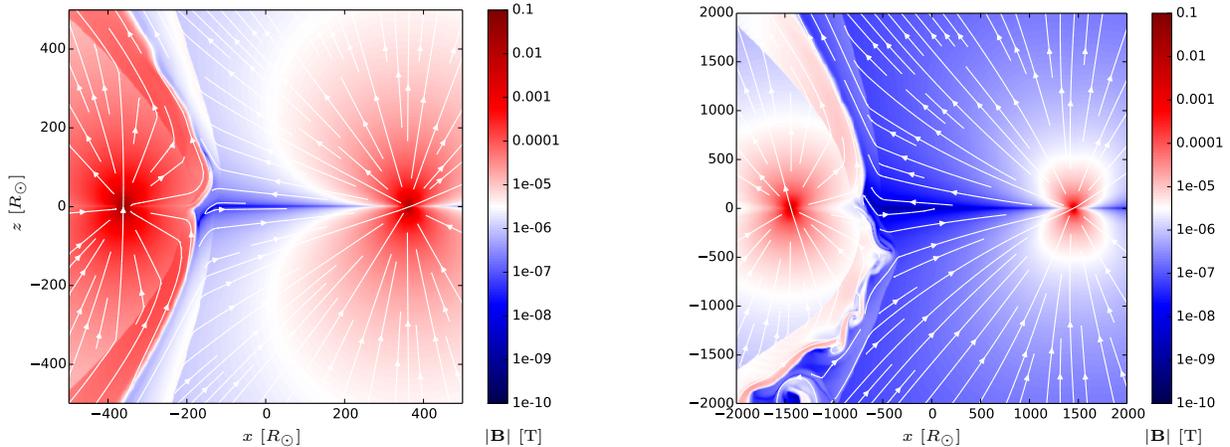
\begin{figure}[ht]
	\setlength{\unitlength}{0.00034\textwidth}
	\begin{picture}(1396,1100)(-100,-100)
	\put(-50,450){\tiny\rotatebox{90}{$z$ [$R_{\sun}$]}}
	\put(510,-50){\tiny$x$ [$R_{\sun}$]}
	\put(1090,-50){\tiny$|\vec{B}|$ [T]}
	\includegraphics[height=1000\unitlength]{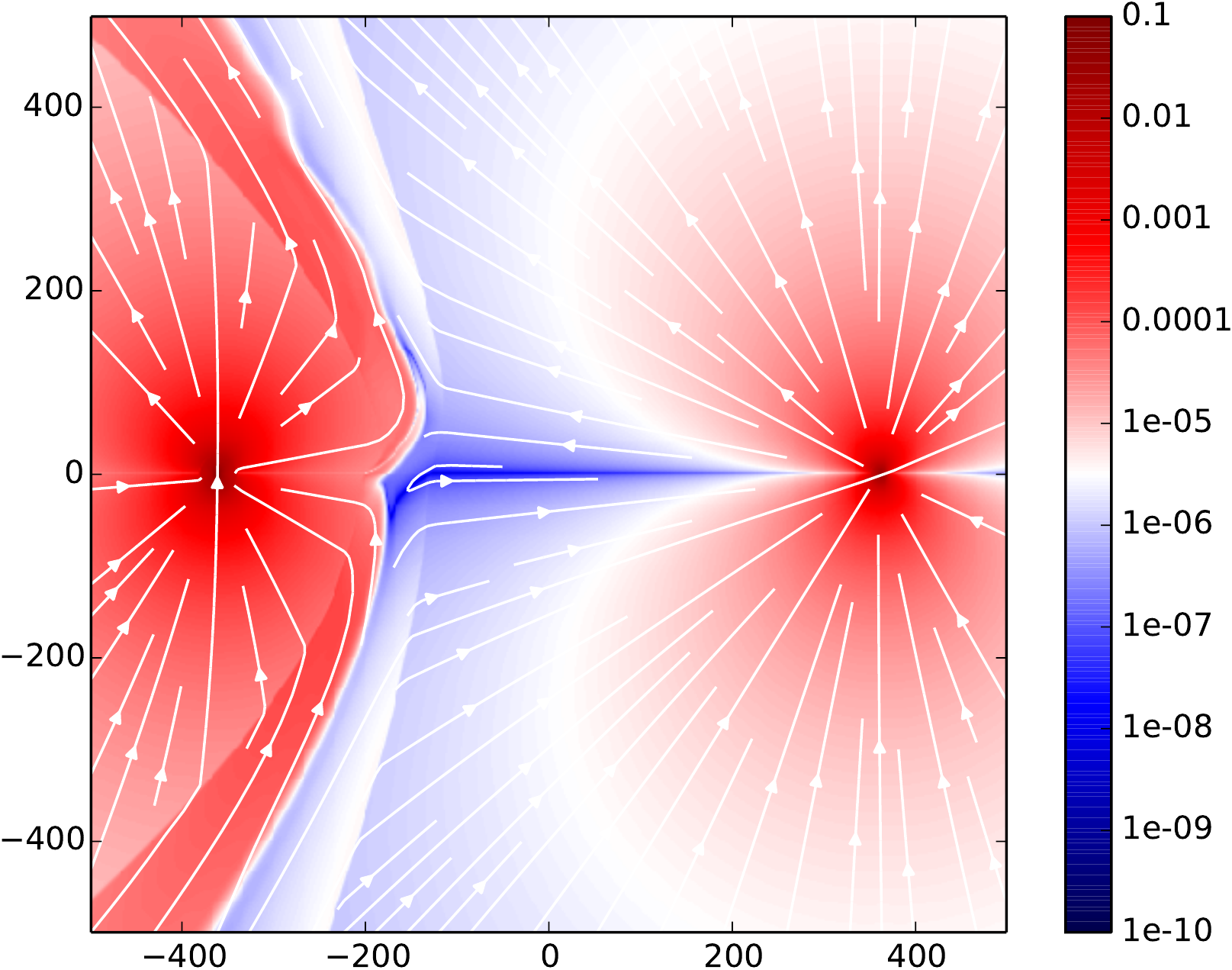}
	\end{picture}
	\hfill
	\begin{picture}(1296,1100)(0,-100)
	\put(510,-50){\tiny$x$ [$R_{\sun}$]}
	\put(1090,-50){\tiny$|\vec{B}|$ [T]}
	\includegraphics[height=992\unitlength]{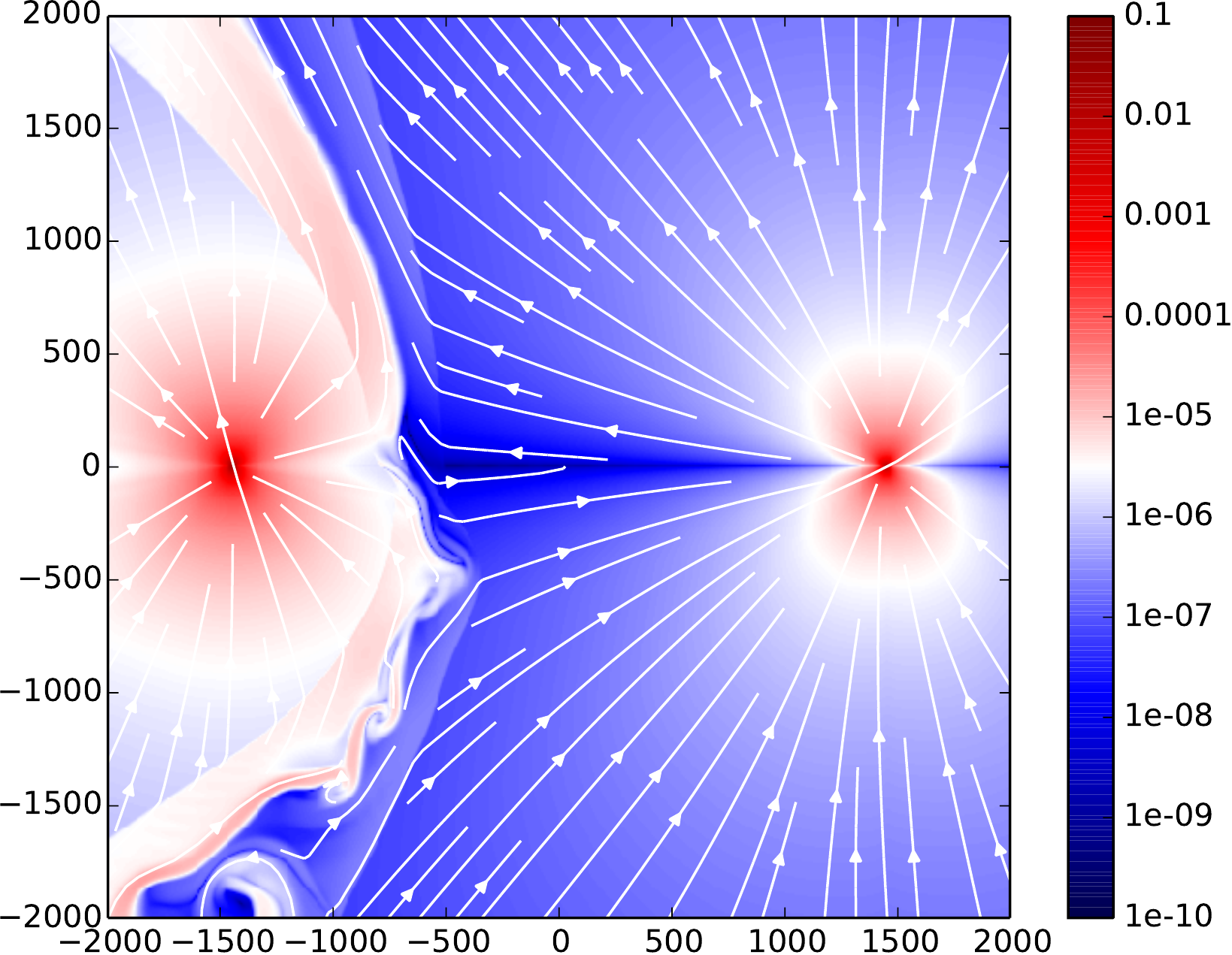}
	\end{picture}
	\caption{\label{FigCWBModelC}Results for the magnetized CWB simulation for setups A1 and A3 with stellar separations of 720\,$R_{\sun}$ (left) and 2880\,$R_{\sun}$ (right). Here, the absolute value of the magnetic induction is shown in the $xz$-plane together with the vector direction.}
\end{figure}
The observations for the time evolution and for the structure of the WCR remain valid also for the simulations with different stellar separations. This is illustrated in Fig. \ref{FigCWBModelC}, where we show results for model A1 and A3 with stellar separations of 720\,$R_{\sun}$ and 2880\,$R_{\sun}$, respectively. In both cases the contact discontinuity in the WCR remains unstable in the direction of the flipped nose structure. 
Expectedly, the magnetic-field strength within the WCR decreases with increasing stellar separation. The Alfv\'en speed, however, is very similar in all cases.
 Additionally, the region, where the shock is quasi parallel around the line of centers between the stars becomes larger with growing stellar separation. Implications of this are further discussed in Sec. \ref{SecAccel}.


The impact of turbulence increases with growing stellar separation despite the use of the same number of grid points for models A1, A2, and A3.  For the largest stellar separation -- model A3 -- the turbulence becomes sufficiently strong to also have a distinct impact on the shocks (see Fig. \ref{FigCWBModelC} on the right).
The increasing impact of turbulence with larger stellar separation can be understood by a close investigation of the evolution of KH-driven turbulence in our numerical code. In a numerical model the evolution of turbulence in the WCR is determined by three competing effects: first, turbulence is driven by the KH instability, for which the growth rate $\Gamma_{KH} \propto \lambda^{-1}$, where $\lambda$ is the wavelength of the instability. Secondly, fluctuations are damped by numerical viscosity. Thirdly, whether turbulent fluctuations can be observed also depends on the time available for the growth of the fluctuations before the fluctuations are advected out of the the numerical domain.

We closely investigated the growth of the KH instability in the initial phase leading to the tilting of the nose structure. The growth of the instability is fastest in the region close to the line of centers between both stars, because there the change in flow speed $\Delta u$ when passing the contact discontinuity is largest, with $\Gamma_{KH} \propto \Delta u$. As discussed in the previous section, this is related to the outward bulge of the WCR caused by the presence of the magnetic field.
Beyond that the growth rate of the KH instability decreases nearly by an order of magnitude. Therefore, fluctuations of a certain wavelength only occur in the simulations when the growth timescale is significantly shorter than the time it takes for the fluctuations to be advected out of the strongly unstable region.

For our standard setup -- model A2 -- damping by numerical viscosity is on the same order as $\Gamma_{KH}$ for fluctuations with $\lambda \simeq 50 R_{\sun}$ \citep[Here, we estimated the numerical damping rate by simulations of a decaying shear flow as discussed in][]{RyuGoodman1994ApJ422_269}.
 Therefore, only longer-wavelength fluctuations can efficiently be driven by the KH instability. Due to $\Gamma_{KH} \propto \lambda^{-1}$ long wavelength fluctuations grow ever more slowly, rendering the growth rate for $\lambda > 200 R_{\sun}$ too slow to lead to significant disturbances in the available time.

This also means that for models A1 and A3 fluctuations with $\lambda > 100 R_{\sun}$ and with $\lambda > 400 R_{\sun}$, respectively, are suppressed because of the finite time available for the growth of the fluctuations. Numerical dissipation approximately goes like $\Gamma_{damp} \propto \lambda^{-1} N^{-2}$ in our simulations where $N$ is the number of cells covering a length scale $\lambda$. This also reveals why the large-scale simulations are more turbulent: for instance we cover a $\lambda = 200 R_{\sun}$ region in model A3 with the same number of cells as a $\lambda = 100 R_{\sun}$ region in model A2, but due to the larger spatial extent the damping rate is smaller by a factor of 2 in model A3. Thus, a larger range of wavelengths is unstable in model A3, which is evident in Fig. \ref{FigCWBModelC}.

This situation would be more severe if all simulations would have been done with the same cell size. In this case, fluctuations in model A1 would be suppressed entirely, while model A3 would have shown a very wide range of unstable wavelengths with the driving most efficient at smallest scales due to $\Gamma_{KH} \propto \lambda^{-1}$. This also implies that future higher-resolution simulations will reveal a wider range of turbulence.



\subsection{Impact of Inclined Dipole Axes}
\begin{figure}[ht!]
\begin{center}
	\setlength{\unitlength}{0.00034\textwidth}
	\begin{picture}(820,1000)(100,-100)
	\includegraphics[trim=3.6cm 0cm 0cm 0cm, clip=true,height=950\unitlength]{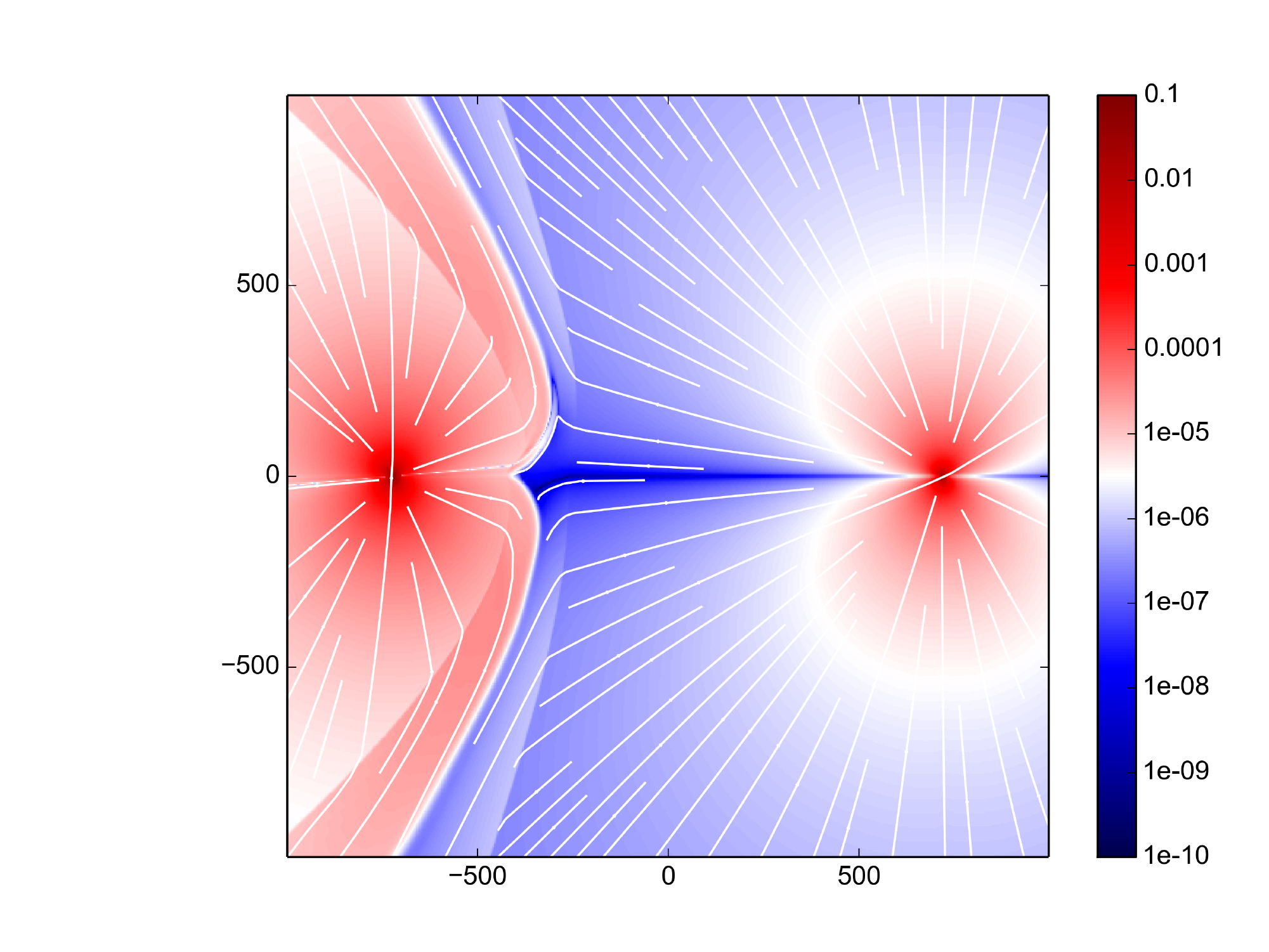}
	\put(-1100,400){\tiny\rotatebox{90}{$z$ [$R_{\sun}$]}}
	\put(-600,20){\tiny$x$ [$R_{\sun}$]}
	\end{picture}
	\begin{picture}(820,1000)(100,-100)
	\includegraphics[trim=3.6cm 0cm 0cm 0cm, clip=true,height=950\unitlength]{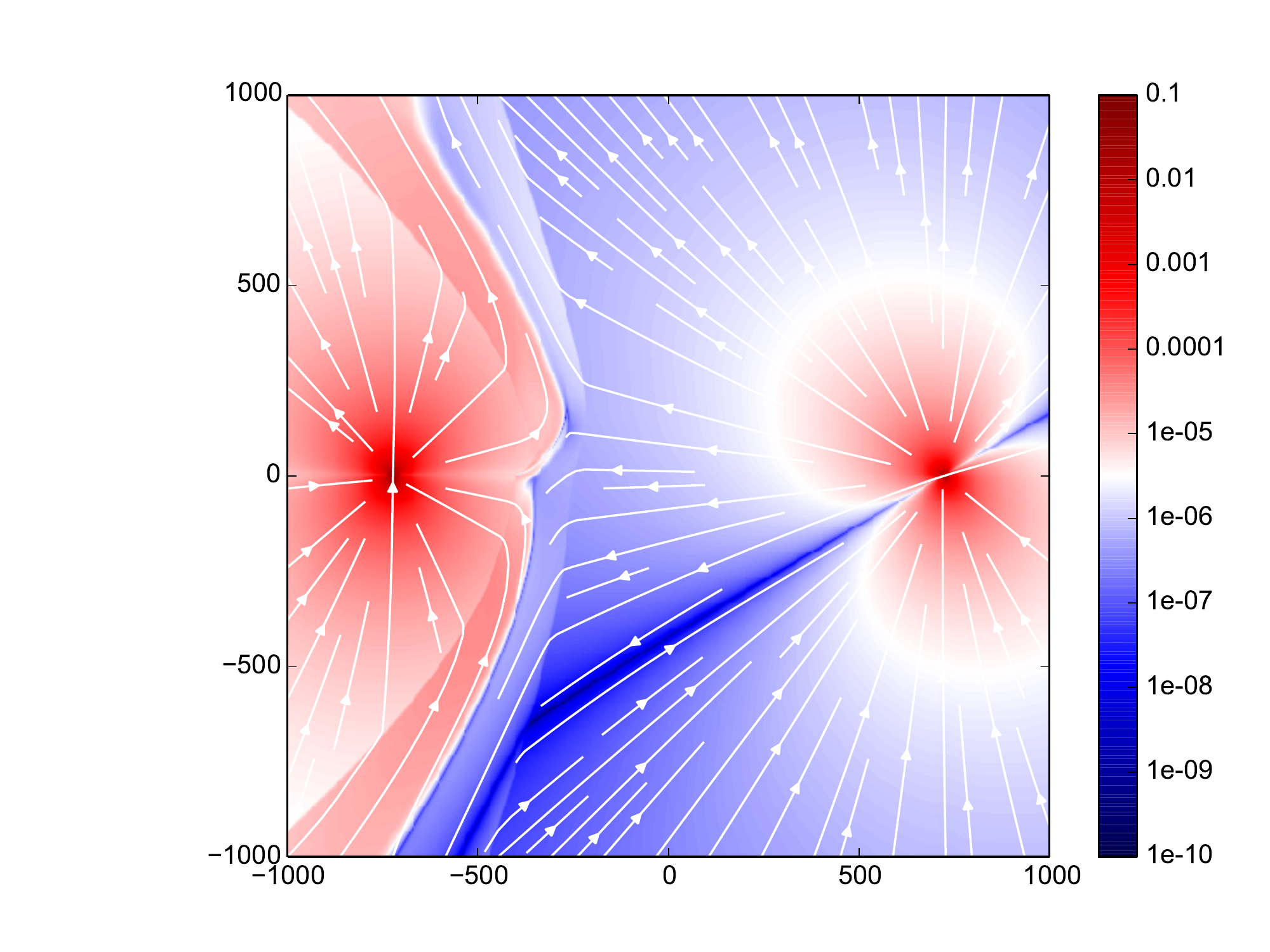}
	\put(-600,20){\tiny$x$ [$R_{\sun}$]}
	\end{picture}
	\begin{picture}(980,1000)(100,-100)
	\includegraphics[trim=3.6cm 0cm 0cm 0cm, clip=true,height=950\unitlength]{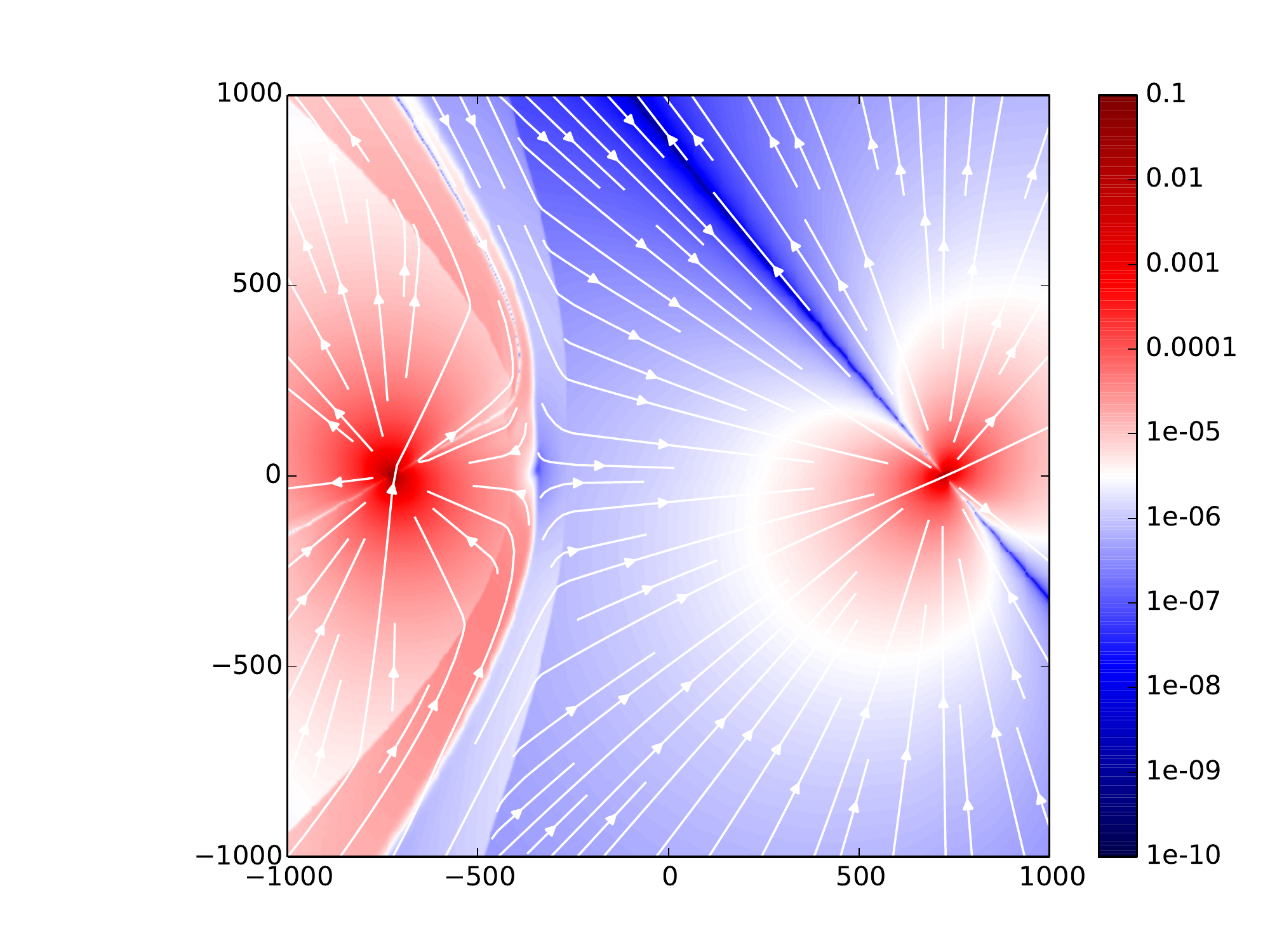}
	\put(-600,20){\tiny$x$ [$R_{\sun}$]}
	\put(-200,20){\tiny$|\vec{B}|$ [T]}
	\end{picture}
	
	\vspace{-0.3cm}
		\begin{picture}(820,1000)(100,-100)
	\includegraphics[trim=3.6cm 0cm 0cm 0cm, clip=true,height=950\unitlength]{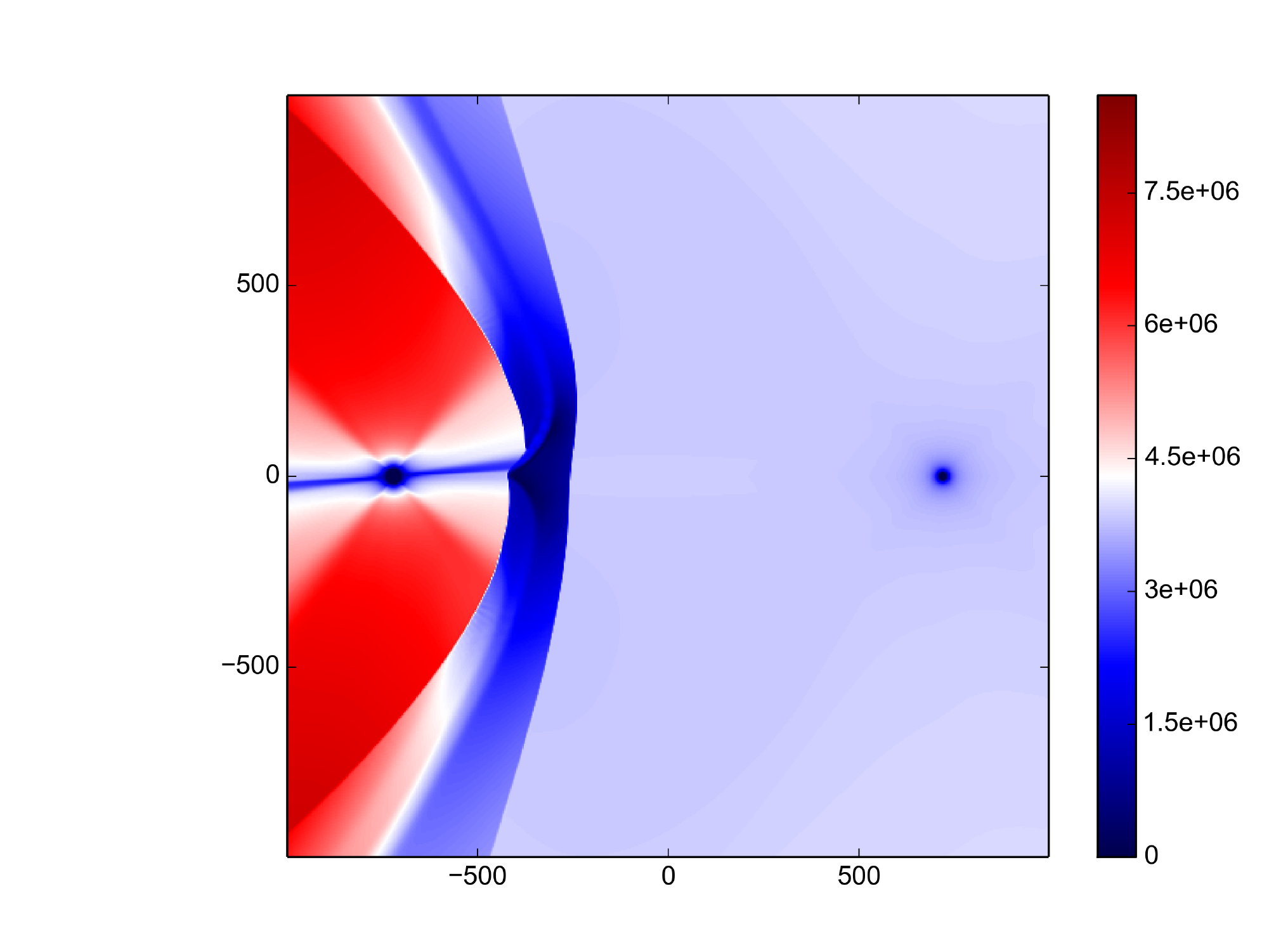}
	\put(-1100,400){\tiny\rotatebox{90}{$z$ [$R_{\sun}$]}}
	\put(-600,20){\tiny$x$ [$R_{\sun}$]}
	\end{picture}
	\begin{picture}(820,1000)(100,-100)
	\includegraphics[trim=3.6cm 0cm 0cm 0cm, clip=true,height=950\unitlength]{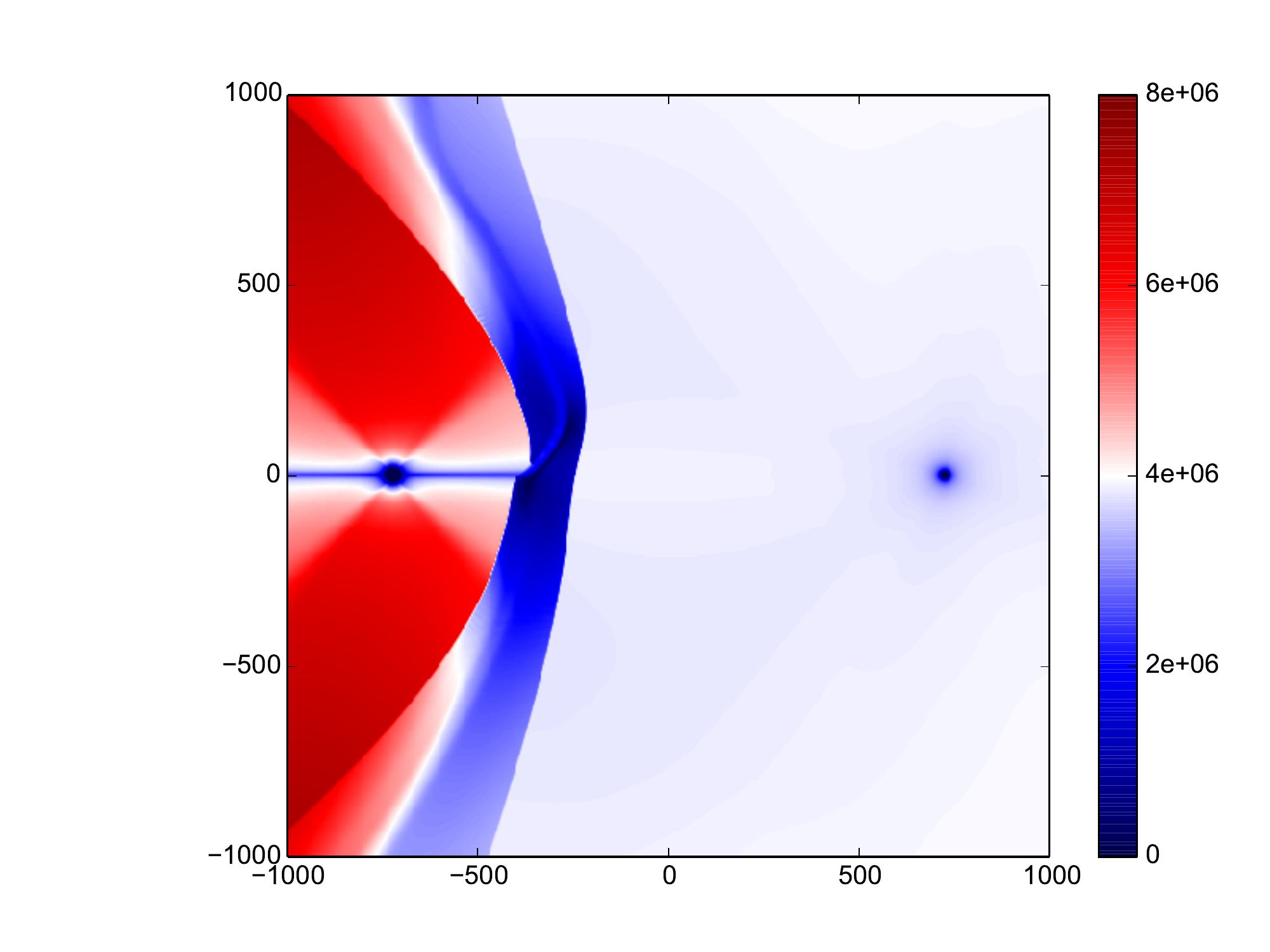}
	\put(-600,20){\tiny$x$ [$R_{\sun}$]}
	\end{picture}
	\begin{picture}(980,1000)(100,-100)
	\includegraphics[trim=3.6cm 0cm 0cm 0cm, clip=true,height=950\unitlength]{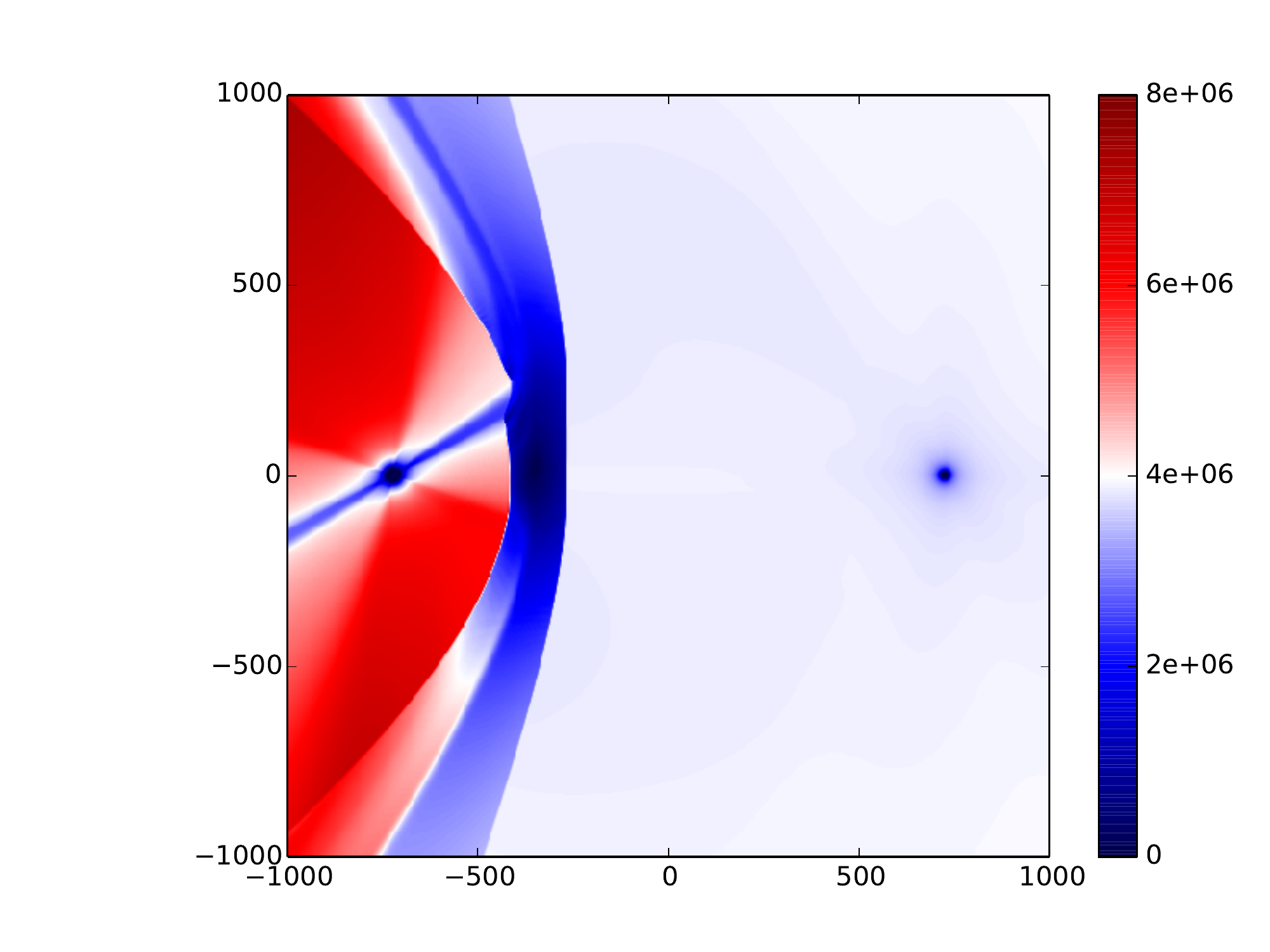}
	\put(-600,20){\tiny$x$ [$R_{\sun}$]}
	\put(-230,20){\tiny$|\vec{v}|$ [m\,s$^{-1}$]}
	\end{picture}
	\end{center}
	\vspace{-1.5cm}
		\caption{\label{vartilt}Results for the magnetized CWB simulation for setups  B1, B2, and B3 for the magnetic induction (top) and the velocity (bottom) in the $xz$-plane for varying orientation of the stellar magnetic dipole fields with field strengths of $0.01$\,T. From left to right the tilt $\delta$ of the dipole axis with respect to the standard orientation changes as follows. Left plot: $\delta_\mathrm{B}=5^\circ$, $\delta_\mathrm{WR}=0^\circ$, middle plot: $\delta_\mathrm{B}=0^\circ$, $\delta_\mathrm{WR}=30^\circ$, right plot: $\delta_\mathrm{B}=30^\circ$, $\delta_\mathrm{WR}=-40^\circ$.
In each case the absolute value is shown. For the magnetic induction we also show the vector direction.}
\end{figure}
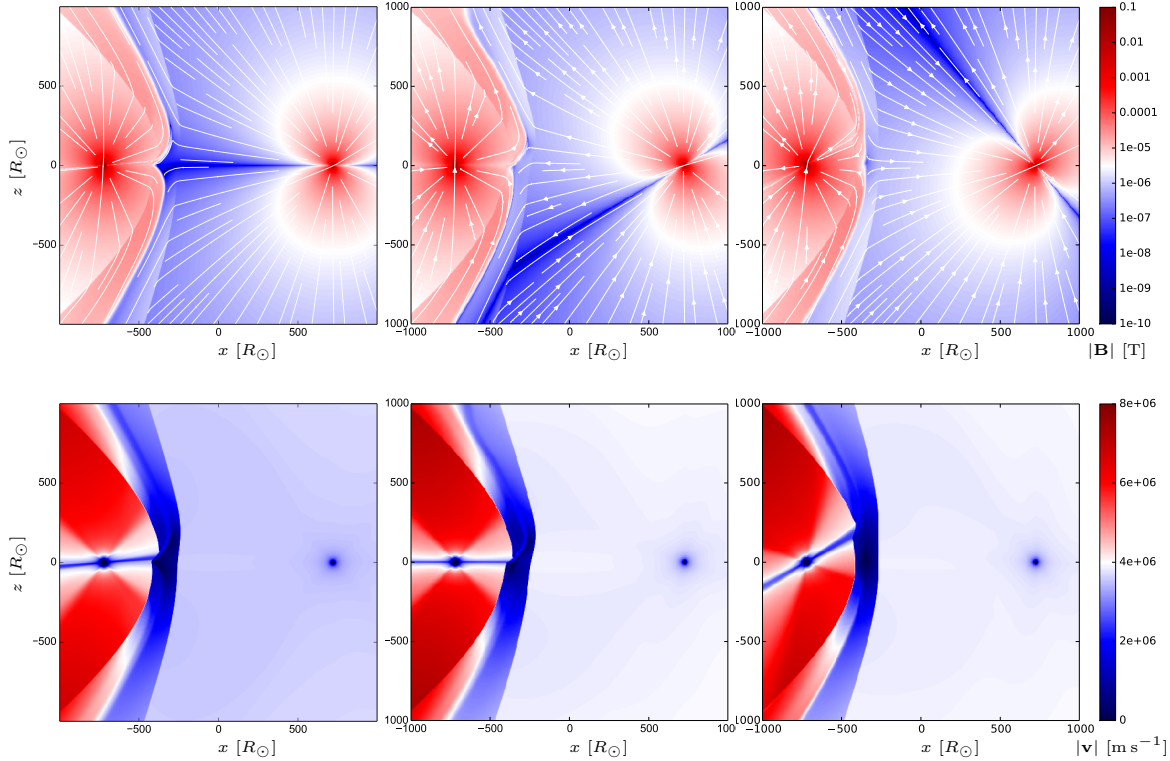

A perpendicular orientation of the line of centers (connecting the stars) and the magnetic field axes (as it has been assumed in our simulations so far), is merely a special case amongst a wider field of possible configurations. In Figure \ref{vartilt} we explore what happens when the magnetic dipole axes of the stars are inclined with respect to the $z$-direction, within the $xz$-plane. The polar magnetic fields are kept constant at $0.01$\,T.

The first column shows results for model B1, in which the magnetic field of the B star is inclined by an angle $\delta_B=5^\circ$. The WCR converges to a smooth state without noticeable instabilities along the contact discontinuity. Due to the initial asymmetry in this case, no nose-like feature emerges, not even as a transitionaory state. The gap at the apex of the WCR is wider and deeper than in the case of dipole axes along the $z$-direction. With the absence of the nose-like structure, the contact discontinuity also does not become turbulent in these simulations. This shows that even such a small inclination of the dipole axis is enough to cause a very different structure of the WCR. Here, the region of higher mass-loss rate is shifted sufficiently far from the equatorial region that instead of a forming nose-like structure the region of higher mass loss is advected upwards without forming an equatorial peak.
Without this peak-like displacement of the WCR the relative velocity of the flow on both sides of the contact discontinuity near the line of centres becomes much smaller. Therefore, the growth rate of the KH instability decreases, thus explaining the absence of turbulence. In this case a turbulent WCR can in this case only be recovered by higher-resolution simulations.

For simulation B2 displayed in the central column of Fig. \ref{vartilt}, the dipole axis of the WR star was inclined by $\delta_B=30^\circ$ with the dipole axis of the B star along the $z$-direction. Comparing this to the case of Figure \ref{FigCWBModelB}, no substantial effect of this inclined dipole of the WR star can be seen -- except that now the nose flipped by chance towards negative $z$ direction. The absence of a larger effect on the WCR is not surprising since a polar magnetic field of $0.01$\,T is still too small to cause noticeable anisotropies in the wind of the WR star. 
Accordingly, the observed instability is not due to some magnetic interaction of the current sheets of both stars but is rather a consequence of the higher radial mass flux in the equatorial region of the B star.


The dipole fields of both stars have been inclined in simulation B3 displayed in the third column of Fig. \ref{vartilt}. They are inclined by $\delta_\mathrm{B}=30^\circ$ and $\delta_\mathrm{WR}=-40^\circ$. It is predominately the inclination of the B star's magnetic field which causes the noticeable difference to the simulation without inclination. Like in setup B1 with the small inclination of the B star's dipole axis (left column in Fig. \ref{vartilt}), the nose structure and the ensuing turbulence do not appear in this case. This shows that the emergence of the nose structure and the effects of the turbulence investigated in the previous sections are related to a special situation where the region of higher mass-loss rate coincides with the direction to the other star. The structure of the WCR is very sensitive to the specific orientation of the stellar dipoles.

\subsection{Variation in Polar Magnetic Field Strength}

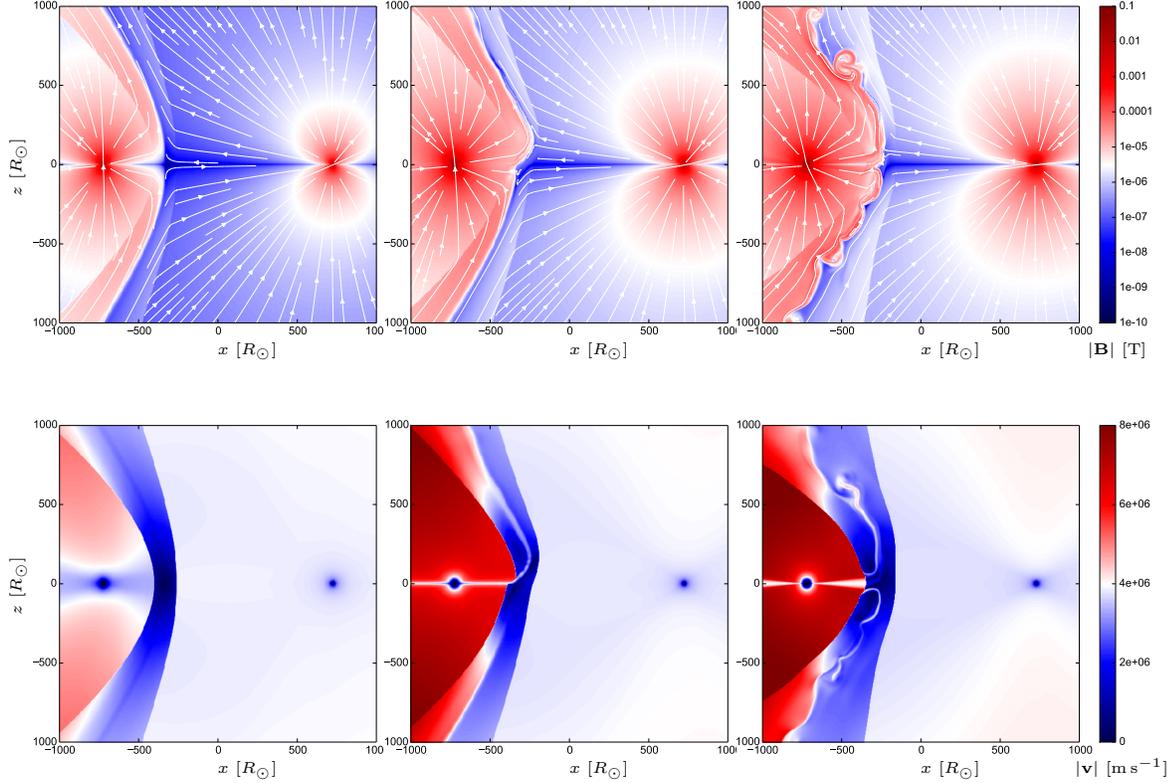
\begin{figure}[t!]
\begin{center}
	\setlength{\unitlength}{0.00034\textwidth}
	\begin{picture}(820,1000)(100,-100)
	\includegraphics[trim=3.6cm 0cm 0cm 0cm, clip=true,height=950\unitlength]{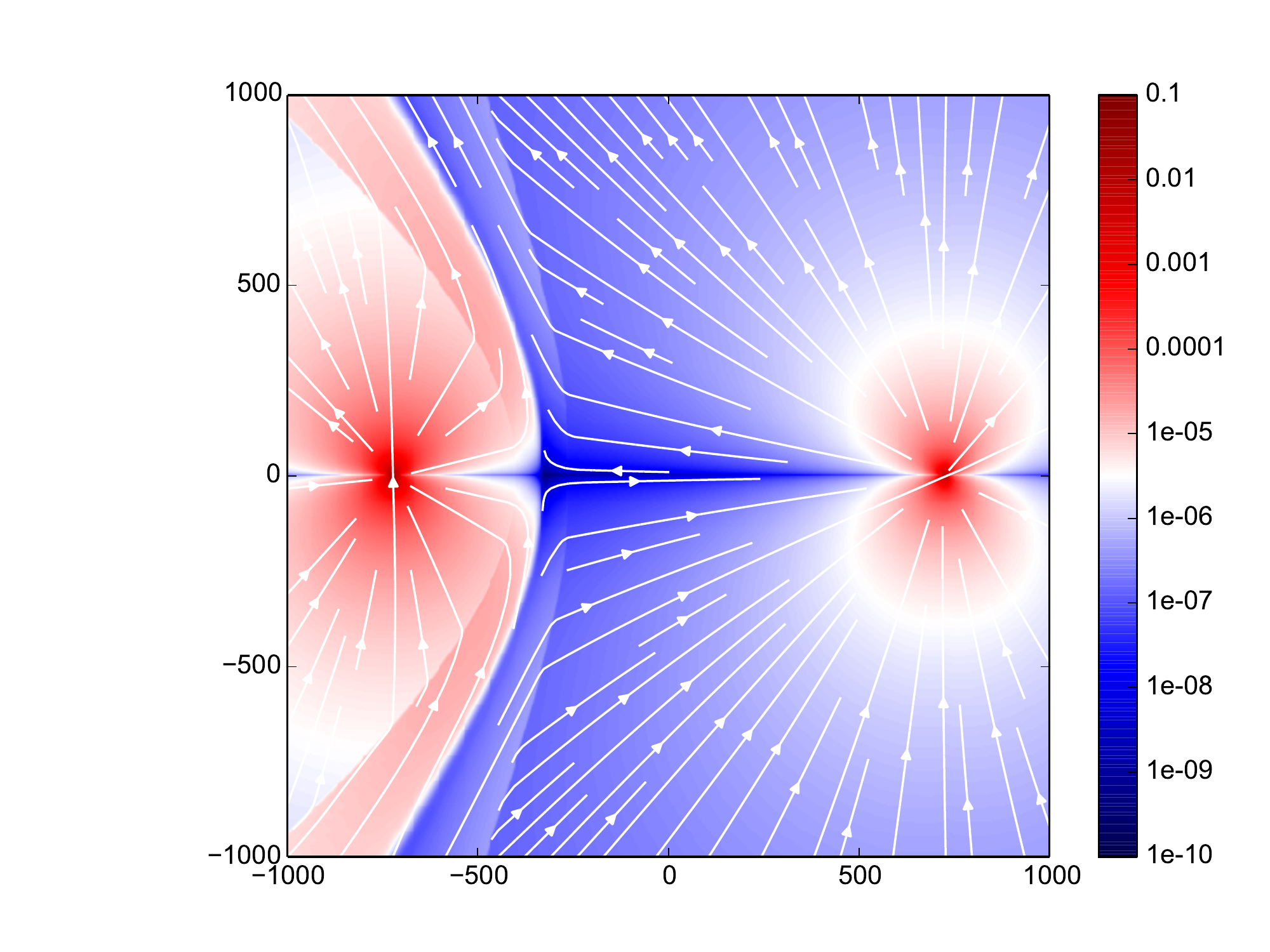}
	\put(-1100,400){\tiny\rotatebox{90}{$z$ [$R_{\sun}$]}}
	\put(-600,20){\tiny$x$ [$R_{\sun}$]}
	\end{picture}
	\begin{picture}(820,1000)(100,-100)
	\includegraphics[trim=3.6cm 0cm 0cm 0cm, clip=true,height=950\unitlength]{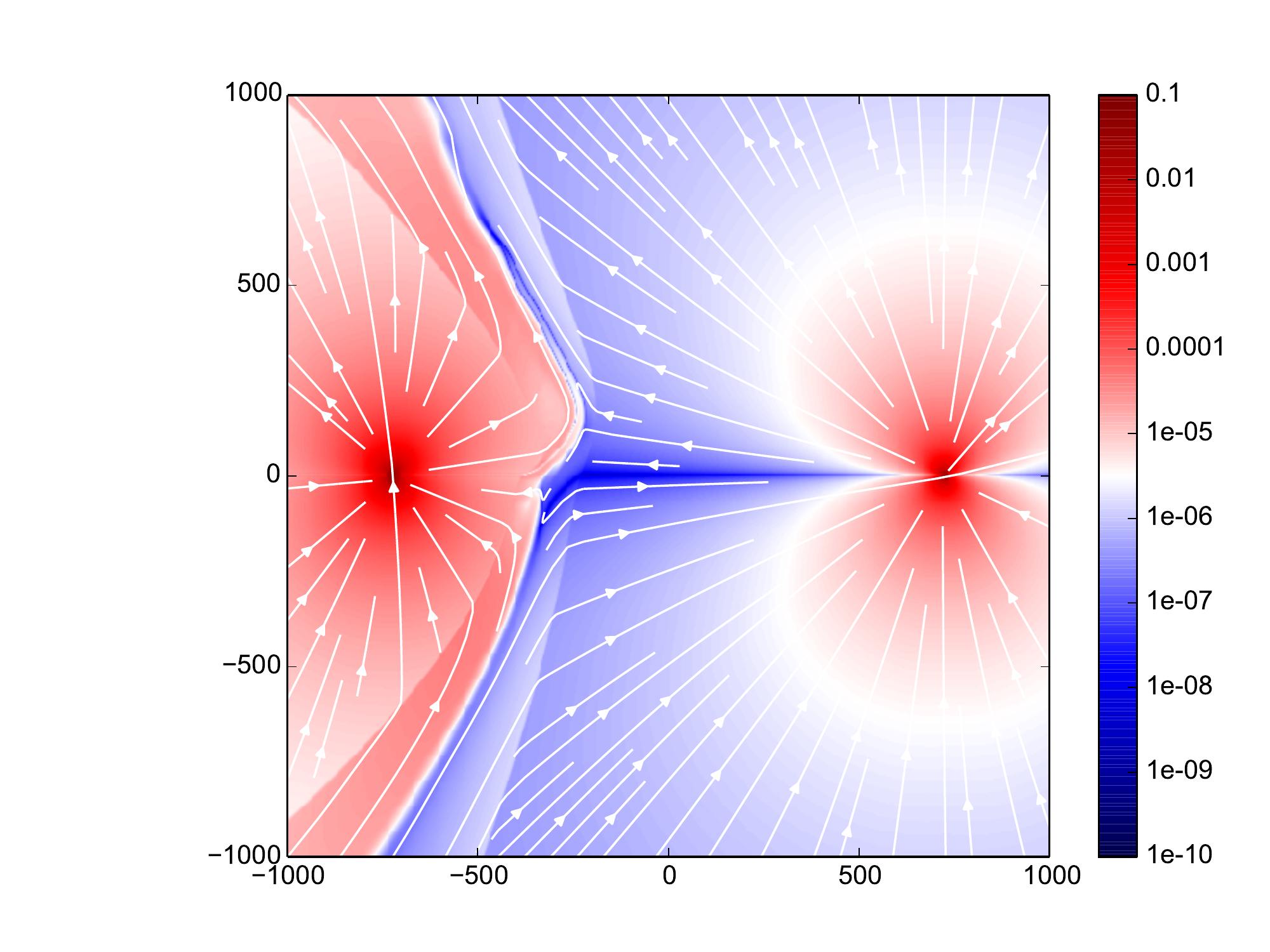}
	\put(-600,20){\tiny$x$ [$R_{\sun}$]}
	\end{picture}
	\begin{picture}(980,1000)(100,-100)
	\includegraphics[trim=3.6cm 0cm 0cm 0cm, clip=true,height=950\unitlength]{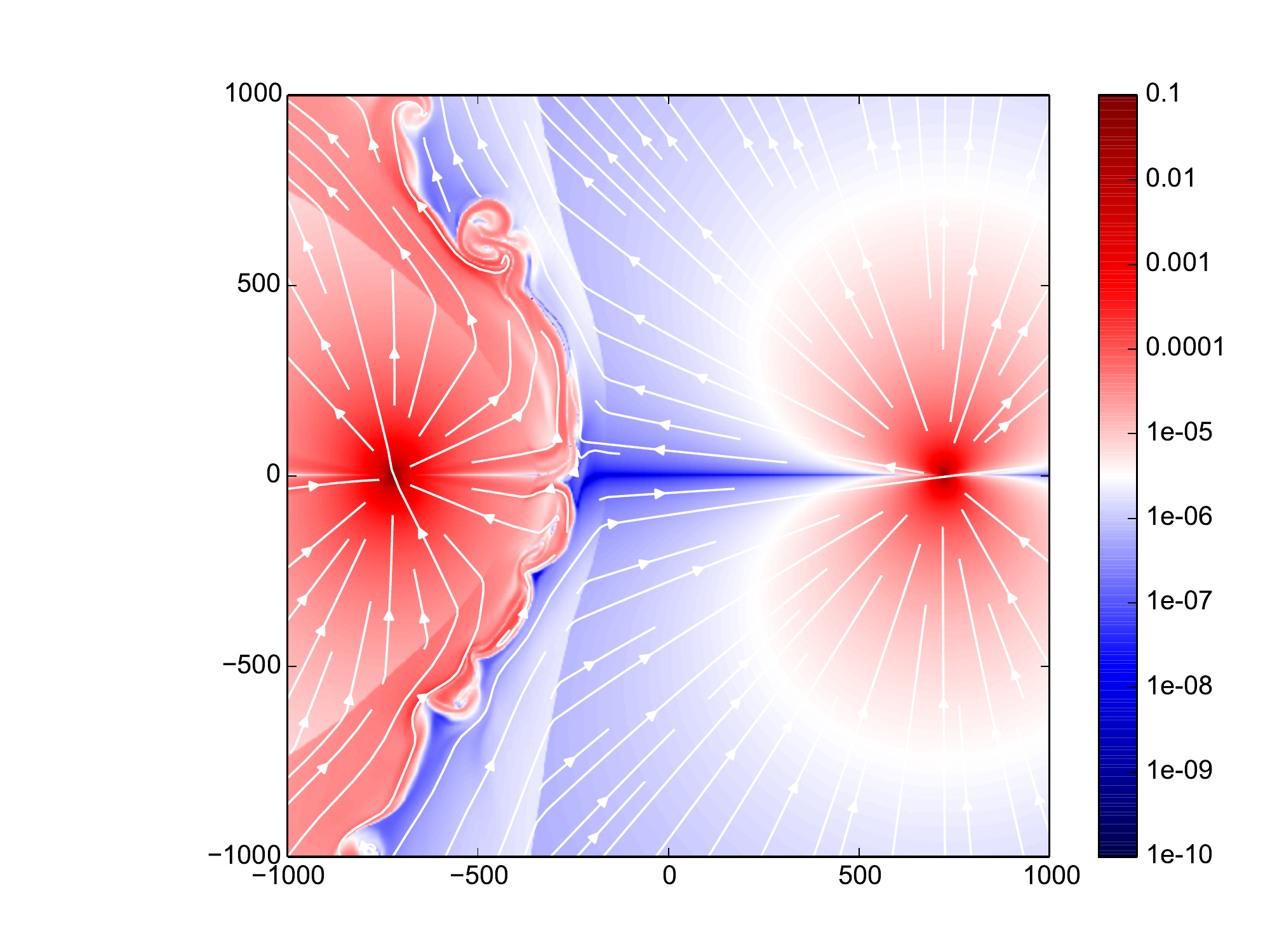}
	\put(-600,20){\tiny$x$ [$R_{\sun}$]}
	\put(-200,20){\tiny$|\vec{B}|$ [T]}
	\end{picture}
	\vspace{-0.3cm}
		\begin{picture}(820,1000)(100,-100)
	\includegraphics[trim=3.6cm 0cm 0cm 0cm, clip=true,height=950\unitlength]{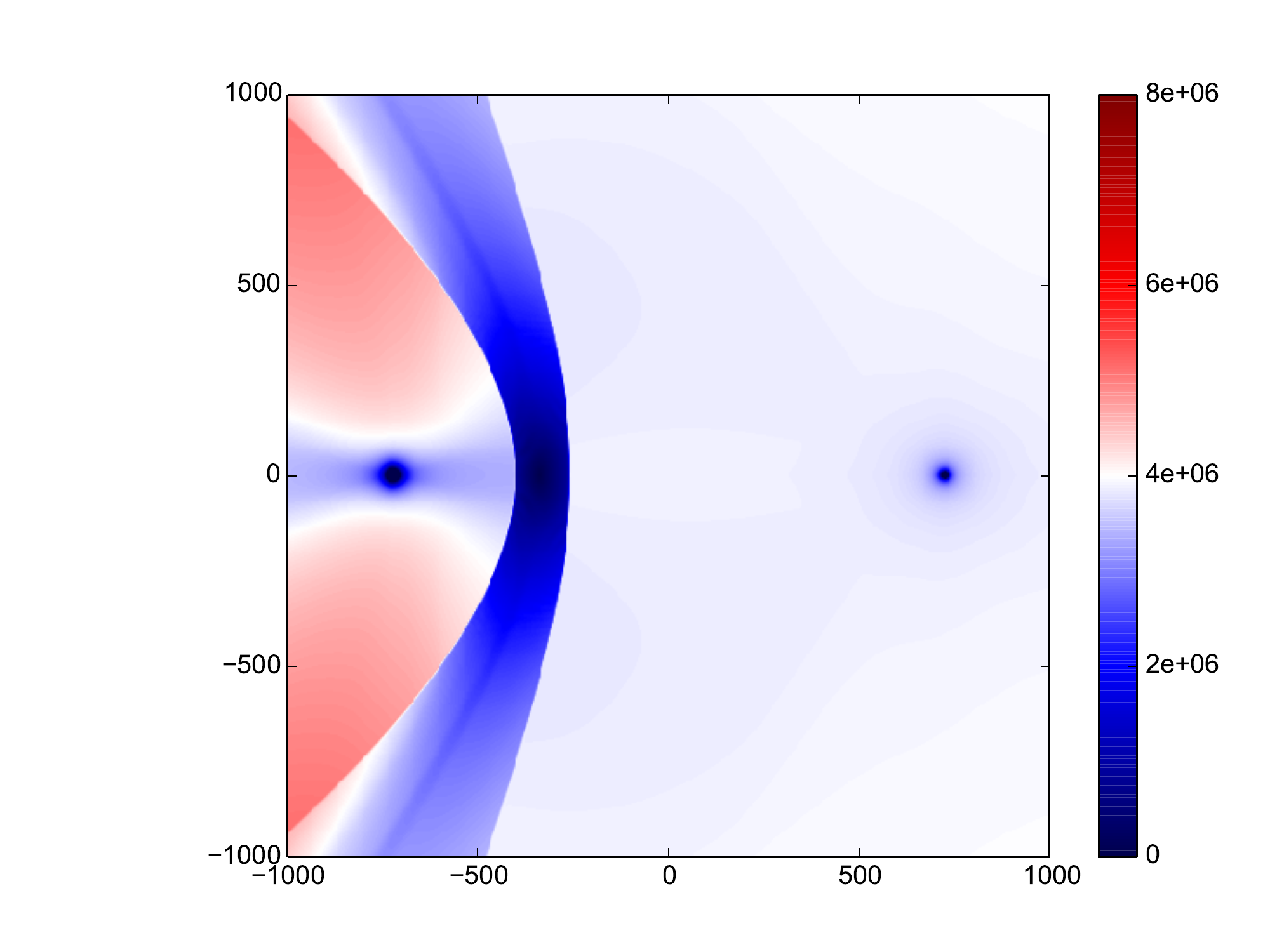}
	\put(-1100,400){\tiny\rotatebox{90}{$z$ [$R_{\sun}$]}}
	\put(-600,20){\tiny$x$ [$R_{\sun}$]}
	\end{picture}
	\begin{picture}(820,1000)(100,-100)
	\includegraphics[trim=3.6cm 0cm 0cm 0cm, clip=true,height=950\unitlength]{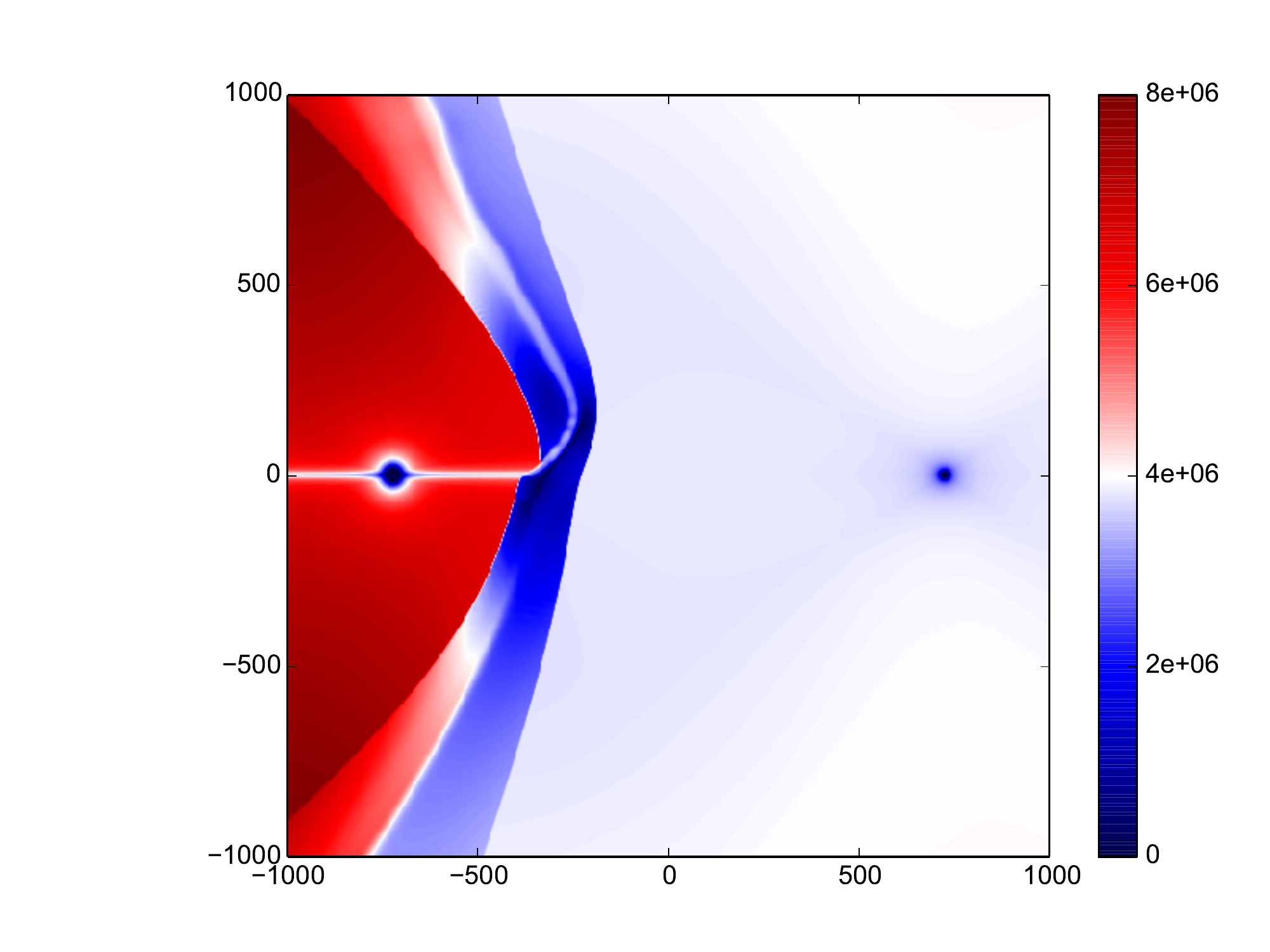}
	\put(-600,20){\tiny$x$ [$R_{\sun}$]}
	\end{picture}
	\begin{picture}(980,1000)(100,-100)
	\includegraphics[trim=3.6cm 0cm 0cm 0cm, clip=true,height=950\unitlength]{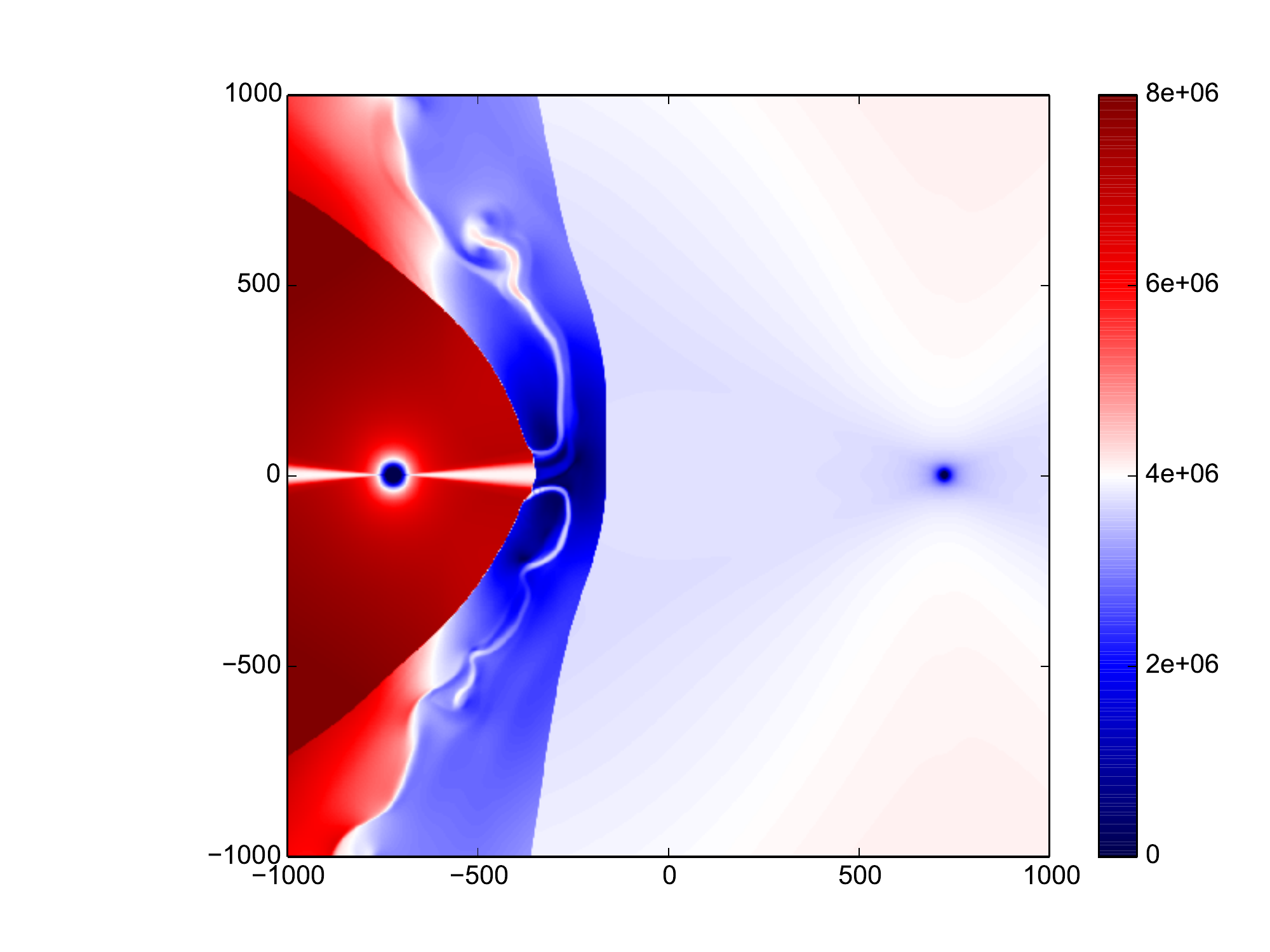}
	\put(-600,20){\tiny$x$ [$R_{\sun}$]}
	\put(-230,20){\tiny$|\vec{v}|$ [m\,s$^{-1}$]}
	\end{picture}
	\end{center}
	\vspace{-1cm}

		\caption{\label{varB}Results for the magnetized CWB simulation for setups C1, C2, and C3 for the magnetic induction (top) and the velocity (bottom) in the $xz$-plane for varying polar magnetic field strengths. From left to right the polar magnetic field strength of both stars was chosen as $0.005$, $0.015$ and $0.02$\,T. In each case the absolute value is shown. For the magnetic induction we also show the vector direction.}
\end{figure}

The distinct impact of the stellar magnetic dipole field on the structure and properties of the WCR as discussed above raises the question how this changes with varying magnetic field strength. Figure \ref{varB} shows our results for simulations C1, C2, and C3 in which the polar magnetic field strength of both stars is $0.005$\,T, $0.015$\,T, and $0.02$\,T. These lead to magnetic confinement parameters for the B star's wind of $\eta = 0.048, 0.432$, and $0.768$, respectively. Simulations for the former two have been performed with a resolution of $512^3$. Model C3 requires higher spatial resolution, owing to the very thin region of enhanced mass-loss rate at the magnetic equator. It was calculated on a grid of dimensions $1024^2\times 128$.

Comparing the respective results to each other, as well as to the previously discussed case of $\eta=0.192$ for the B star's wind reveals intriguing insights into the delicate dependence of the WCR on the magnetic field environment as  shown by the following discussion of the structure of the WCR.

For $\eta = 0.048$ the shape of the resulting WCR very much resembles our findings in earlier purely hydrodynamic simulations \citep[see][Fig. 1]{ReitbergerEtAl2014ApJ782_96}. The magnetic field of the B star and the anisotropies in density and velocity it causes are yet too small to lead to a noticeable deformation of the WCR. As detailed above, this changes for $\eta = 0.192$ and even more so for $\eta=0.432$ where the well collimated beam of high density material in the B wind also causes the emergence of the nose-like feature as in Figure \ref{FigModelBvartime}.

For the case of a very high polar magnetic field of $0.02$\,T the nose feature only appears for the higher-resolution grid. This is because the region of higher mass-loss rate is even more sharply collimated around the equator in this case. In this case the strong collimation starts nearer to the star than in the cases of a weaker magnetic field. Actually for the standard near-star simulation setup the region of increased mass-loss rate is still underresolved as was to be expected since the magnetic confinement factor is near unity. Therefore, the near-star simulation was done with double resolution in the $\theta$-direction. In that case we find the region of increased mass-loss rate confined to a region $\sim$2.5\,$R_{\sun}$ above and below the equator at the outer edge of the fixed region of the B star at 40\,$R_{\sun}$.

Keeping in mind that the standard setup for the colliding-wind binary simulation uses 512 cells to cover the range from -1000\,$R_{\sun}$ to 1000 \,$R_{\sun}$ leading to a cell size of $\sim$3.9\,$R_{\sun}$ shows why in this case also the colliding-wind binary simulation had to be done using higher spatial resolution, as noted above. In this case we, again, find the nose-like feature leading to a turbulent contact discontinuity. This also shows that setups with $\eta>1$ for one of the stars would require considerably higher-resolution simulations.

Another feature that becomes apparent in Figure \ref{varB} is the significant difference in the downstream speed of the shocked gases in the WCR for higher field strengths. There is not much difference between downstream speeds on both sides of the WCR in model C1 with $B_{\text{polar}} = 0.005$\,T.  For $B_{\text{polar}} = 0.015$\,T and $B_{\text{polar}} = 0.02$\,T, however, the increased stellar wind velocity towards the magnetic poles of the B star leads to distinctively higher velocities in the outer wings of the WCR towards the B star -- about a factor of 2 higher than in the wing towards the WR star. This reflects the impact of the stellar magnetic field on the wind acceleration for the B star. As discussed in section Sec. \ref{SecSimSingle} both the latitudinal structure and the terminal velocity  of the stellar wind are altered in presence of magnetic fields. In consequence, velocities on both sides of the shock discontinuity vary drastically. Change in structure of the WCR is therefore a manifestation of the presence of magnetic fields in realistic CWB simulations.


Finally, we observe that for the case of $0.02$\,T, wind anisotropy is now also clearly visible in the undisturbed wind around the WR star (lower right panel in Fig. \ref{varB}).

\subsection{Outlook: Impact on Particle Acceleration}
\label{SecAccel}
The variation in the strength of the magnetic field within the WCR -- especially for models with different stellar separations -- in conjunction with the shock geometry relative to the magnetic field also has important implications for particle acceleration. In this context the distinction between parallel and perpendicular shocks is made, referring to the relative direction of the shock normal and the magnetic field direction.

While most of the classic work on diffusive shock acceleration focused on particle acceleration at parallel shocks \citep[see, e.g.,][]{AxfordEtAl1977ICRC11_132,Krymskii1977DoSSR234_1306,Bell1978MNRAS182_147,BlandfordOstriker1978ApJ221L_29} -- with the notable exception of \citet{Schatzmann1963AnAp26_234} -- more recent studies also address the acceleration capability of perpendicular shocks \citep[see, e.g.][]{Ostrowski1988MNRAS233_257, EllisonEtAl1995ApJ453_873, TakamotoKirk2015ApJ809_29}. Apart from that, numerical studies show that also the dynamical evolution of the shock geometry may become important for the particle acceleration efficiency \citep[see, e.g.,][]{SandroosVainio2006AnA455_685,PomoellVainioKissmann2011ASTRA7_387}. While this indicates that our understanding of diffusive shock acceleration theory is not complete yet, it also underlines the importance of the shock geometry for particle acceleration.

The previously discussed models allow an analysis of this shock geometry for both shocks of the WCR. We investigated the relative direction of the magnetic field and the shock normal upstream and downstream of each of the shocks for models A1, A2, and A3. In our simulations the shocks are sufficiently far from the stars that the magnetic field approximately points along the flow direction in the upstream direction of each shock. Thus, the region showing a quasi-parallel shock grows with stellar separation.

In our analysis, we identify the shock position from the velocity gradient: the centre cell of the shock is given as the position, where the velocity gradient is smallest. 
 The shock normal is identified from the temperature gradient. The magnetic field angle is then computed as the local angle between the computed shock normal direction and the magnetic field at a position of $3 \Delta x$ along the direction of the shock normal, both in the upstream and downstream direction.

\begin{figure}[ht]
	\setlength{\unitlength}{0.00042\textwidth}
	\begin{picture}(1100,860)(-100,-100)
	\put(-50,350){\tiny\rotatebox{90}{angle}}
	\put(450,-50){\tiny$h$ [$R_{\sun}$]}
	\includegraphics[width=1000\unitlength]{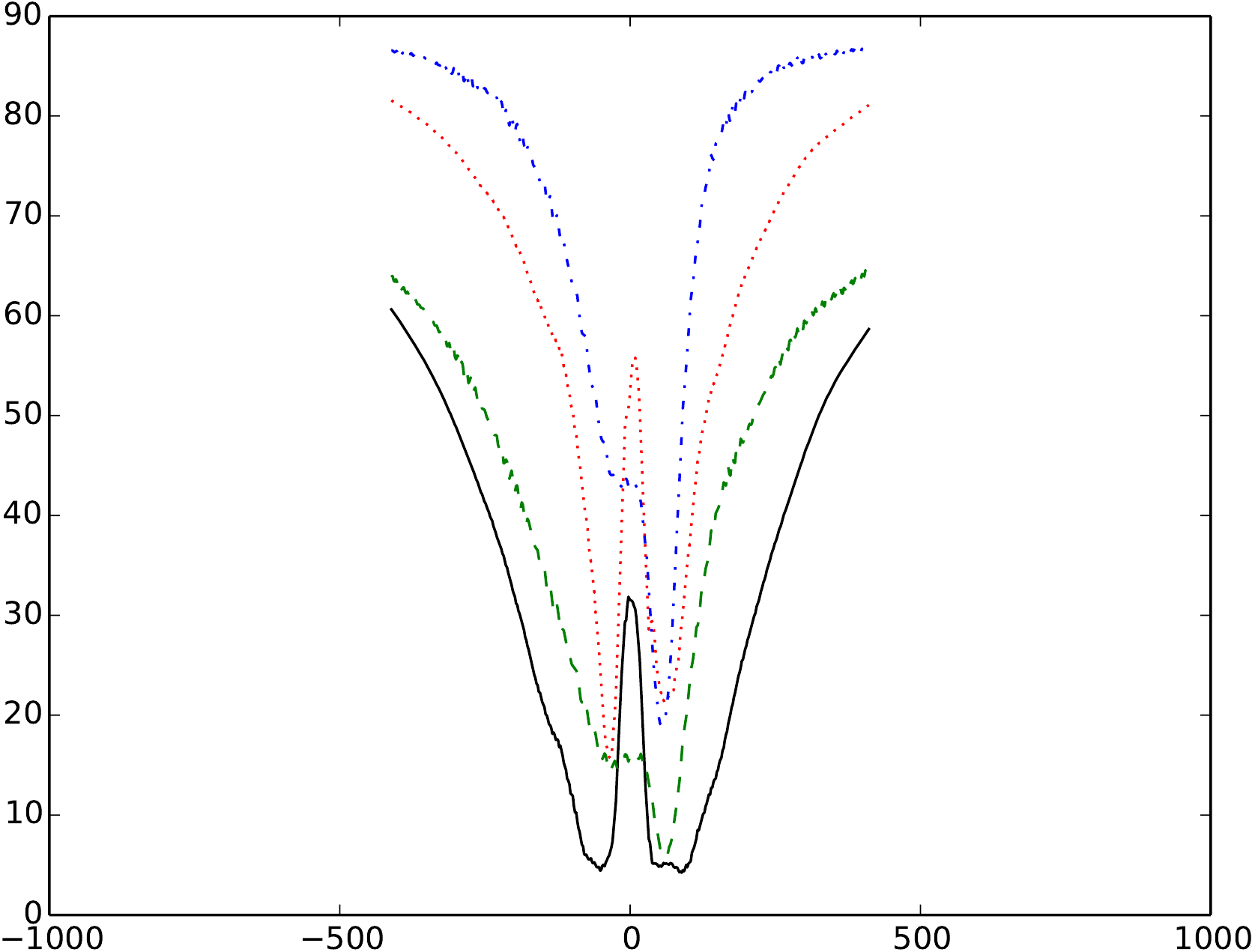}
	\end{picture}
	\hfill
	\begin{picture}(1000,860)(0,-100)
	\put(450,-50){\tiny$h$ [$R_{\sun}$]}
	\includegraphics[width=1000\unitlength]{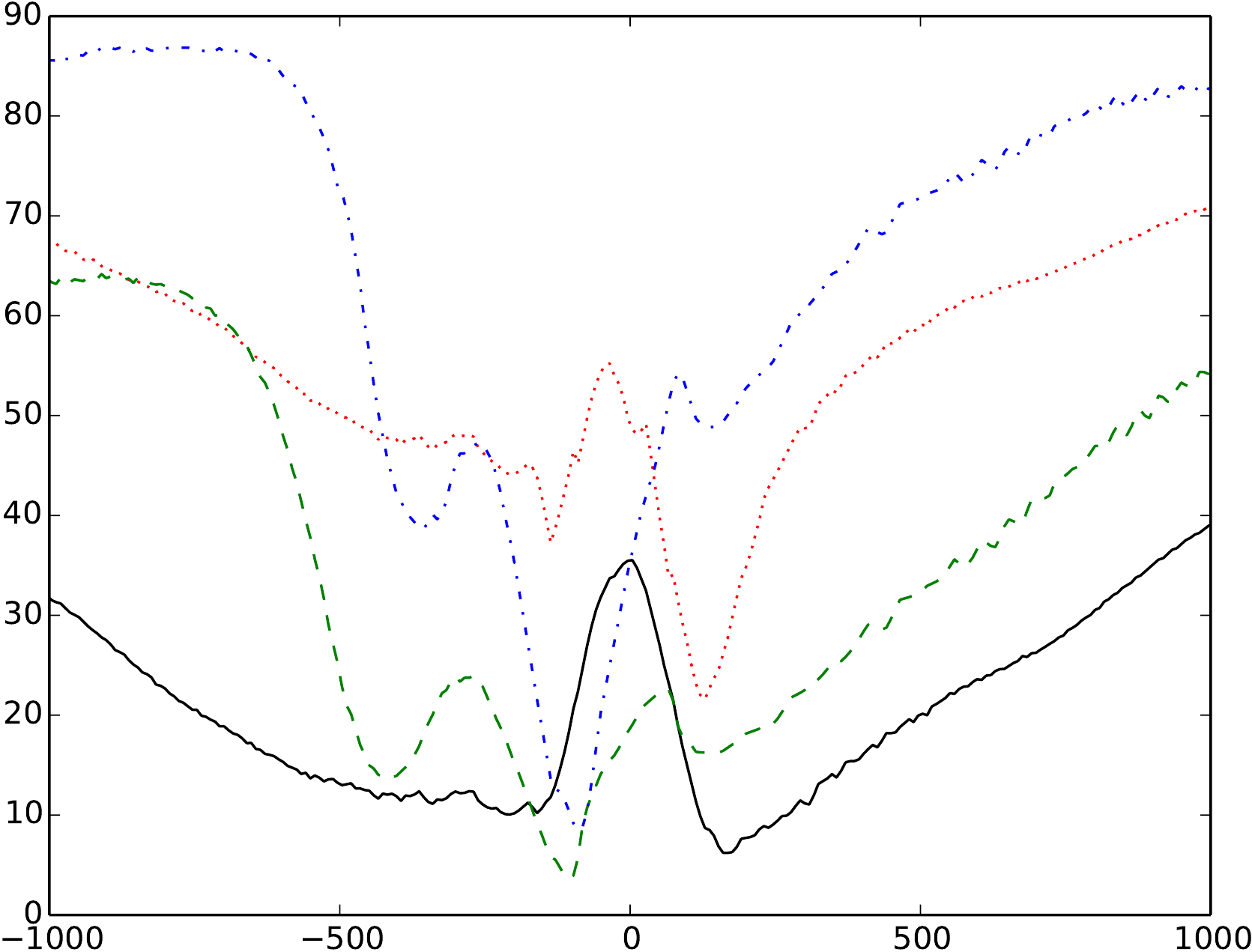}
	\end{picture}
	\caption{\label{FigMagAngle}Angle between magnetic field direction and shock normal for models with a stellar separation of $d=720$\,$R_{\Sun}$ (left) and $d=2880$\,$R_{\Sun}$ (right) as a function of the distance from the $xy$-plane $h$. For both models we show results up- and downstream of the B star's shock (black solid line and red dotted line) and up- and downstream of the WR star's shock (green dashed line and blue dash-dotted line). Some smoothing was applied to diminish fluctuations caused by the interpolation.}
\end{figure}

Results are shown in Fig. \ref{FigMagAngle} for models A1 and A3, where the angles were computed both up- and downstream of the B and the WR star's shock, i.e. the shocks located closer to the respective star. The region, where the shock is quasi parallel is larger for the shock induced by the B star's wind in all models, because of the curvature of the WCR. Due to the increase in the perpendicular component of the magnetic field downstream of the shock, the related magnetic-field angle is always larger than the upstream one.

An exception to this is the peak near the $xy$-plane visible for the B star's shock. This is related to the flipped nose structure that results in a shock normal that deviates significantly from the $x$ direction in the $xy$-plane. An additional analysis of the setup with inclined dipole axes for both stars, shows that the overall structure remains rather similar to the initial setup. Only the peak is shifted according to the direction of the magnetic equator. Thus, the overall effect of a tilted dipole on the acceleration efficiency should be rather small. In that context the suppression of the turbulence within the WCR will probably prove to be more important.


As was to be expected, models with a bigger stellar separation lead to a correspondingly larger region, where the shocks are quasi-parallel. This possible effect on the acceleration efficiency of each shock comes in addition to the difference in possible energy losses for the particles that are reduced with increasing separation of the stars. In this regard a correct description is of particular importance for the synchrotron losses. A consistent description of the magnetic field is required especially for the energy losses and the acceleration efficiency of the energetic particles in a CWB.

For higher resolution simulations the WCR can be expected to show stronger turbulence at small spatial scales. As is already hinted at by model A3, this can lead to localized and dynamical changes in the direction of the shock fronts bounding the WCR. Especially, this dynamical behavior of the shocks might have relevant implications for particle acceleration and limits the application for particle acceleration in post--processing. In such a case particle acceleration should be simulated together with the fluid-dynamical simulation of such systems, e.g., in \citet{ReitbergerEtAl2014ApJ782_96}.


\section{Conclusion}
\label{SecConclusion}
In this study we detailed a numerical method for the investigation of colliding-wind binary systems that are subject to sufficiently strong stellar magnetic fields such that the structure of the stellar wind outflows is clearly affected by the presence of these fields. Our method is a multi-step process, where we first make sure that the wind acceleration near the stellar surface is consistently simulated by performing near-star simulations for each star individually. These simulations are then injected into a large-scale simulation of the colliding-wind binary system.

For the magnetic field strengths investigated in this paper we found a significant impact on the structure of the WCR when the polar field strengths of the B star yields a magnetic confinement parameter $\eta > 0.1$. For stronger fields a nose-like structure emerges at the WCR connected to the higher mass-loss rate near the magnetic equator of the B star. This nose structure -- that becomes ever sharper with increasing strength of the surface magnetic field of the B star -- is unstable and decays, leaving one side of the WCR in a turbulent state. Apparently, this affects mainly the contact discontinuity, leaving the shock fronts mostly undisturbed -- at least for the simulations with stellar separation smaller than 2880\,$R_{\Sun}$. This disturbance only becomes relevant, when the dipole axis of the B-star is normal to the line of centers between the stars. Thus, the specific setups of the dipole axes are very important for the structure of the WCR. Together with the dynamics of such a system, this might lead to WCRs becoming more turbulent during short episodes of the stellar orbit. 

In this study we only investigated a limited amount of configurations. Especially for magnetic fields near or even exceeding a magnetic confinement parameter $\eta\ge 1$, considerably higher spatial resolution will be necessary to be able to study the impact of the ever thinner region with increased mass loss around the magnetic equator.
Thus, the present study is only a starting point showing that the impact of the presence of the magnetic field is very relevant and should be taken into account in future investigations.

Apart from the impact of the magnetic field on the structure of the WCR also the direction of the magnetic field relative to the shock fronts is relevant for particle acceleration. We show the first analysis of the obliqueness of the shocks in the WCR. The most important parameter in this context is the stellar separation, where in our model an increasing separation leads to a larger region where the shock is quasi parallel. However, stronger turbulence in the WCR can also have an effect on the shock-fronts possibly leading to very localized and dynamical changes in shock obliqueness. This might only be revealed in higher-resolution simulations. Such an analysis of the shock obliqueness allows for a consideration of different acceleration efficiencies in future simulations of particle acceleration in magnetized colliding-wind binaries. This also shows the importance of a correct treatment of the magnetic field in the interpretation of observed high-energy emission patterns from these binary systems.

\acknowledgments

This work was supported by the Austrian Ministry of Science BMWF as part of the UniInfrastrukturprogramm of the Focal Point Scientific Computing at the University of Innsbruck. Emanuele Grimaldo and Anita Reimer acknowledge financial support from the Austrian Science Fund (FWF), project P 24926-N27.


\bibliography{$HOME/LaTeX/Bibliographies/GalacticCR,$HOME/LaTeX/Bibliographies/numerics,$HOME/LaTeX/Bibliographies/books,$HOME/LaTeX/Bibliographies/ExtraGalacticCR,$HOME/LaTeX/Bibliographies/pubrk,$HOME/LaTeX/Bibliographies/Discs,$HOME/LaTeX/Bibliographies/solarwind,$HOME/LaTeX/Bibliographies/GammmaBinaries}

\end{document}